\documentclass[11pt]{article}
\usepackage{epsfig}
\usepackage{persdefs}
\newlength{\dinwidth}
\newlength{\dinmargin}
\newlength{\figwidth}
\newcommand\gH[1]{\gammh{H}{#1}}
\newcommand\gK[1]{\gammh{K}{#1}}
\newcommand\gstarH[1]{\gstar{H}{#1}}
\newcommand\gstarK[1]{\gstar{K}{#1}}

\newcommand\suml{\sum_{l\in\Z^d}}

\newcommand\Hl{\Hh(l)}

\newcommand\epsKtext{\prod_{\nu\in K}             (-1)^{l_\nu}}
\newcommand\psih{ \hat{\psi}}
\newcommand\psibh{\hat{\psib}}

\newcommand\varphib{\bar{\varphi}}

\newcommand\phixH{  \varphi (x,H) }

\renewcommand{\Z}{Z \!\!\! Z}
\renewcommand{\R}{{\kern+.25em\sf{R}\kern-.78em\sf{I} \kern+.78em\kern-.25em}}
\newcommand\fb{ \bar{f}}
\newcommand\ft{ \tilde{f}}
\newcommand\Ft{ \tilde{F}}
\makeatletter
\@addtoreset{equation}{section}
\makeatother

\begin{document}

\vspace*{1cm}
\begin{center}
{\LARGE The Schwinger Model \\
\ \\
with Perfect Staggered Fermions}

\vspace*{1cm}

W. Bietenholz$^{\rm ~a}$ and H. Dilger$^{\rm ~b}$
\footnote{Supported in part by Deutsches Arbeitsamt}
\vspace*{7mm}

$^{\rm a}$ 
%HLRZ c/o Forschungszentrum J\"{u}lich \\
%D-52425 J\"{u}lich, Germany \\
%{\it and} \\
NORDITA \\
Blegdamsvej 17 \\
DK-2100 Copenhagen \O, Denmark \\

\vspace*{7mm}

$^{\rm b}$ Institute for Theoretical Physics I \\
WWU  M\"{u}nster \\
Wilhelm-Klemm Str. 9 \\
D-48149 M\"{u}nster, Germany

\vspace*{1cm}

Preprint \ NORDITA--98/67 HE

\end{center}

\vspace*{1cm}

We construct and test a quasi-perfect lattice action for 
staggered fermions.
The construction starts from free fermions, where
we suggest a new blocking scheme, which
leads to excellent locality of the perfect action.
An adequate truncation preserves a high quality
of the free action.
An Abelian gauge field is inserted in $d=2$ by effectively tuning
the couplings to a few short-ranged lattice paths,
based on the behavior of topological zero modes.
We simulate the Schwinger model with this action, 
applying a new variant of Hybrid Monte Carlo, 
which damps the computational
overhead due to the non-standard couplings.
We obtain a tiny ``pion'' mass down to very small $\beta$,
while the ``$\eta$'' mass follows very closely the prediction of
asymptotic scaling.
The observation that even short-ranged quasi-perfect actions
can yield strong improvement is most relevant in view of QCD.

\newpage

%%%%%%%%%%%%%%%%%%%%%%%%%%%%%%%%%%%%%%%%%%%%%%%%%%%%%%%%%%%%%%%%%%%%%%%%%%%
\section{Introduction}

Most QCD simulations so far
have been performed either using Wilson fermions \cite{Wilfer}
or staggered fermions \cite{KS}. The latter formulation
is especially useful in the chiral limit, because the remnant chiral
symmetry $U(1) \otimes U(1)$ protects the zero fermion mass
from renormalization. As a related virtue, its artifacts due
to the lattice spacing $a$ are only of $O(a^{2})$, whereas
they are of $O(a)$ for Wilson fermions interacting by gauge fields.

It is now widely accepted that
the above lattice actions should be improved, so that
the lattice spacing artifacts are suppressed and coarser
lattices can be used \cite{LAT97}.
There are essentially two improvement strategies in the literature.
In Symanzik's program \cite{Sym} the action is improved
in orders of $a$. For QCD with Wilson fermions
this has been realized on-shell to the first order on the
classical level \cite{SW} and recently also on the
quantum level \cite{Alpha}. Less work has been devoted
to the improvement of staggered fermions, perhaps because
the artifacts in the standard formulation are already smaller.
However, S. Naik has applied Symanzik's program on-shell, where he
improved the free staggered fermion by adding more couplings
along the axes \cite{Naik}, and
the Bielefeld group did the same by adding diagonal couplings
\cite{Bielef}. 
Furthermore, the MILC collaboration achieved a 
reduced pion mass by treating the gauge variable as a 
``fat link'' \cite{MILC}.
Finally, some work on improved operators has been done in
this framework \cite{oper}.

The other promising improvement scheme is non-perturbative
in $a$ and uses renormalization group concepts. 
It has been known for a long time that there are perfect
lattice actions in parameter space, i.e.\ actions without
any cutoff artifacts \cite{WiKo}. More recently, it has been
suggested to approximate them for asymptotically free theories
as ``classically perfect actions'' \cite{HN}, which works very 
well in a number of two dimensional models \cite{HN,GN,Toy,Lang},
and it has also been applied in 4d pure Yang-Mills gauge theory 
\cite{gauge1,gauge2}.

For free or perturbatively interacting fields, perfect actions
can be constructed analytically in momentum space. For Wilson
type fermions this has been carried out to the first order in the
gauge coupling in the Schwinger model \cite{Schwing} and in QCD 
\cite{QuaGlu}. A technique called ``blocking from 
the continuum'' was extremely useful for this purpose. One expresses
all quantities in lattice units after the blocking, and sends the 
blocking factor to infinity. Hence the blocking process starts from
a continuum theory, and it does not need to be iterated in order
to identify a perfect action.

For staggered fermions, a block variable renormalization group
transformation (RGT), which does not
mix the pseudo flavors and which does therefore preserve the
important symmetries, has been suggested in Ref.\ \cite{KMS}.
It requires an odd blocking factor $n$. Iterating the $n=3$
block variable RGT, a fixed point action, i.e.\ a perfect action
at infinite correlation length, has been constructed 
\cite{GN,Dallas}. Also for staggered fermions blocking from the
continuum is applicable \cite{MaiMack}. 
This has been carried out for a general (flavor non-degenerate)
mass term,
revealing the intimate relation to the Dirac-K\"{a}hler fermion 
formulation in the continuum \cite{Dilg},
and also including a suitable treatment of the gauge field
\cite{BBCW}.
Using the generalization to a flavor non-degenerate mass \cite{Dilg}, 
the spectral doublers inherent to the staggered fermion formulation 
might be treated as physical flavors in a QCD simulation.

%For any blocking scheme, the perfect action displays the exact
%continuum scaling behavior.
%In practice, however, the approximations in the determination of
%an applicable quasi-perfect action violate this property to some
%extent. With this respect, the choice of the RGT is very important.

In the present paper, first the procedure of blocking
staggered fermions from the continuum is revisited.
We then discuss the optimization of locality,
in the sense of an extremely fast exponential decay of the
couplings in coordinate space. This property
is crucial for practical applications, because we have to 
truncate the couplings to a small number, which is tractable in
simulations.
Of course the truncation violates the perfectness,
but for excellent locality this violation is not too harmful.
%Moreover, the virtues of staggered fermions
%come into play: the truncation induces errors of $O(a^{2})$
%at most, and the remnant axial symmetry is preserved in the
%truncated perfect action, hence it still excludes
%additive mass renormalization ( the latter causes severe problems
%for truncated perfect fermions of the Wilson type \cite{LAT96,TdG}).

We suggest a new blocking scheme, which we call ``partial decimation''.
It leads to a higher degree of locality
than the usual block average method, i.e.\ to a faster
exponential decay of the couplings in coordinate space. 
We then truncate the couplings to a short range by
means of mixed periodic and anti-periodic boundary conditions,
which are particularly adequate for staggered fermions.
The excellent quality of the truncated perfect action for
free staggered fermions is confirmed by spectral and
thermodynamic considerations.

We then proceed to the two flavor Schwinger model \cite{Schwinger}
as a test case. The Schwinger model
is well-understood from continuum calculations, also in finite volume
\cite{Sachs,AJ,Seiler}. It shares important features with QCD, 
such as asymptotic freedom, confinement and 
topological quantum numbers with corresponding zero-modes of the 
Dirac operator. On the other hand,
one expects other features of higher-dimensional gauge 
theories not to be well-represented by the Schwinger model, due to its 
super-renormalizability.

The fermion gauge vertex is added ``by hand'' to the
truncated perfect free fermion. We first insert the 
$U(1)$ links between fermionic source and sink along
certain shortest lattice paths. In a second step,
we implement ``fat links'' for the paths consisting of
just one link. The staple weight is determined effectively
by minimizing the eigenvalues of the approximate 
topological zero-modes.
%present for non-vanishing topological charge of the 
%gauge field are as close as possible to zero.
For the pure gauge part, we use two actions,
which are perfect resp.\ approximately perfect in $d=2$.
%As pointed out in Ref.\ \cite{QuaGlu}, this can be 
%achieved with single plaquette couplings. 
%Here, it is easily derived using 
%the gauge independent plaquette variables, 
%which are not subject to constraints 
%in contrast to higher dimensional gauge theories. 

Of course, the ad hoc treatment of the vertex
deviates from the systematic
construction of a (classically) perfect action.
However, in the staggered scheme any approximation 
or truncation of perfect action can cause errors of 
$O(a^{2})$ at most, and the remnant axial symmetry protects the mass term 
from renormalization. By contrast, a strong mass renormalization
is a severe problem for quasi-perfect fermions 
of the Wilson type \cite{LAT96,TdG,Kostas,Norbert}.

Naively, the computational effort for simulations increases 
about linearly with the number of couplings in the action.
In addition, physical effects, e.g.\ more exact zero-modes, 
may be a hurdle for the application of an improved action.
We propose a way to damp this increase of computational 
costs, exploiting the freedom to design the Molecular Dynamic steps in the 
Hybrid Monte Carlo algorithm. We use a simplified action there, and the 
full quasi-perfect action in the Metropolis acceptance decision.
We discuss the performance of this method for different values
of $\beta$, the crucial question being the acceptance rate.
This method may also accelerate
simulations with improved actions on parallel machines.
It depends on a sufficiently good simplified version of the improved 
action, which is another plus for staggered fermions.

For quasi-perfect staggered fermions, the dispersion relation and
the scaling behavior of the mass spectrum are good down to 
$\beta $~\raisebox{-.4ex}{$\stackrel{<}{\sim}$}~1.5. 
For these low values of
$\beta$, the masses for the $\pi$- and $\eta$-particle 
demonstrate a good scaling resp.\ asymptotic scaling behavior.
In particular the scaling is even better than the one observed
in Ref.\ \cite{Lang}, which uses truncated perfect Wilson type 
fermions, and 123 independent couplings parameterizing a
classically perfect fermion-gauge vertex. 
Compared to that scheme, a drastically reduced number of
couplings is needed here, and even more in the step to $d=4$.
Our results 
%-- in particular the very small $\pi$ mass --
demonstrate that in fact a relatively modest
number of couplings can yield a very powerful improvement.

A synopsis of this work was presented in Ref.\ \cite{Lat98}.

%%%%%%%%%%%%%%%%%%%%%%%%%%%%%%%%%%%%%%%%%%%%%%%%%%%%%%%%%%%%%%%%%%%%%%%%%%%
\section{The staggered blocking scheme} \label{Scheme}

The construction of perfect actions for free staggered fermions by blocking
from the continu-um has been described in Refs.\ \cite{Dilg,BBCW}. 
We start by briefly reviewing this procedure. 
The staggered blockspin fermions are defined in two steps. First we transform
the $N_f=2^{d/2}$ flavors of continuum Dirac spinors 
$\psi_a^b(x)$ ($a$: spinor
index, $b$: flavor index) into the Dirac-K\"ahler (DK) representation 
given by $\phixH$.
\footnote{For the relation of the DK formulation of continuum 
fermions \cite{EK62} with staggered lattice fermions we 
refer to Ref.\ \cite{BJ}. The relation to the
block spin transformation is discussed in Ref.\ \cite{Dilg}.}
These functions originate from the representation of
inhomogeneous differential forms
\eqa \label{components}
%
%    \Phi \ = \ 
    \sum_H \phixH \, dx^H \ , \ \ 
    dx^H \ = \ dx^{\mu_1} \wedge \dots \wedge dx^{\mu_h} \ ,
\eqb
where $H=\{\mu_1,\dots,\mu_h\},\ \mu_1 < \dots < \mu_h $ is a multi-index.
Transformation and inverse transformation read
\Eqa
        \phixH & = & \frac{1}{\sqrt{N_f}} \sum_{ab} \gstarH{ab} \, 
                       \psi_a^b(x) \ , \ \ \ 
      \gamma^H \ = \ \gamma^{\mu_1} \gamma^{\mu_2} \dots \gamma^{\mu_h} \ ,
                                                          \label{psitophi}\\
   \psi_a^b(x) & = & \frac{1}{\sqrt{N_f}} \sum_H \gH{ab} \, \phixH \ .
                                                            \label{phitopsi}  
\Eqb

Second, we introduce a coarse lattice of double unit spacing
$\Gammab=\{\yb \,|\, \yb_\mu= 2 \nb_\mu\}$, which is a sublattice of
$\Gamma =\{ y  \,|\,  y_\mu=n_\mu \}$, with $\nb_\mu,n_\mu\in\Z$. The fine
lattice points $y$ are uniquely decomposed as ($\muh$ is the unit vector in 
$\mu$--direction)
\eqa
    y \ = \ \yb \, + \, e_H \ , \ \ e_H \ = \ \sum_{\mu\in H} \muh \ .
\eqb 
Thus the multi-index $H(y)$ defines the position of a fine lattice 
point $y$ with respect to the coarse lattice $\Gammab$.
Now the blockspin variables $\Psi (y)$ can be defined as averages of the
component functions $\varphi(x,H(y))$, 
with a normalized weight $\Pi(x\!-\!y)$,
$\int\!dx\,\Pi(x\!-\!y)=1$, which is assumed to be even and
peaked around $x=y$,
\eqa \label{blockspin}
\Psi(y) \ = \ \frac{1}{\sqrt{N_f}} \sum_{ab}   
            \gstar{H(y)}{ab} \, \int dx \ \Pi(x-y) \ \psi_a^b(x) \ .
\eqb
This scheme has been proposed first in Ref.\ \cite{MaiMack}. 
Its peculiarity is that
the staggered block centers depend on the multi-index $H$ of the 
Dirac-K\"ahler component functions $\phixH$. 
Block average (BA) means in this case average over the overlapping lattice
hypercubes $[y]=\{x | -1 \leq (x_\mu-y_\mu ) \leq 1 \}$. 
This scheme is given by
$\Pi = \Pi_{BA}$, $\Pi_{BA}(x) = 2^{-d}$ for $x\in[y]$, 
and $\Pi_{BA}(x) = 0$ otherwise.

For the following calculation we diagonalize the lattice action using the
staggered symmetries. Here it is important that fine lattice shifts are no
symmetry transformations. However, combination with site-dependent sign 
factors gives rise to the non-commuting flavor symmetry transformations
\cite{symmetry}.
Therefore we replace ordinary Fourier transformation by harmonic analysis
with respect to flavor transformations and coarse lattice translations. 
We thus obtain a modified momentum representation which intertwines Fourier
transformation and the transition back from DK fermions to the Dirac basis,
\begin{eqnarray}
   \Psi_a^b(p) & = & \sum_y \, e^{ipy} \, \gammh{H(y)}{ab} \, \Psi(y)
 \ , \nonumber \\
\bar \Psi_a^b(p) & = & \sum_y \, e^{ipy} \, \gammh{H(y)}{ab} \, 
\bar \Psi(y) \ , 
\qquad  p \in \Bc \ = \ ]- \pi /2 , \pi /2 ]^{d}
\label{ytopab}
\end{eqnarray}
($\Bc$ is the Brillouin zone with respect to the coarse lattice).
Inserting \eqref{blockspin} we find
\eqa
\Psi_a^b(p) \ = \ \frac{1}{\sqrt{N_f}} \int \frac{dp'}{\pi^{d}}
\sum_{a'b'} \ \Pi(p') \ 
 \psi_{a'}^{b'}(p') \ \sum_y \ e^{i(p-p')y} \ 
 \gammh{H(y)}{ab} \gstar{H(y)}{a'b'} \ ,     \label{direct}
\eqb
where $\psi_a^b(p), \Pi(p)$ denote the Fourier transforms of $\psi_a^b(x),
\Pi(x)$. The last sum can be re-written as
\Eqa
    \sum_{\yb\in\Gammab}\,e^{iq\yb}\ \sum_K e^{iqe_K} \ \gK{ab}\gstarK{a'b'} 
    & = & \pi^d \suml
\delta(q-\pi l)\ \sum_K \prod_{\mu\in K} (-1)^{l_\mu}
          \ \gK{ab} \gstarK{a'b'} \nonumber\\
    & = & N_f \, \pi^d \suml \delta(q-\pi l) \ \gammh{\Hl}{aa'} 
                                   \gdagg{\Hl}{b'b} \ . 
\Eqb
We have used the orthogonality of the $\gamma$--matrix elements
\eqa \label{sumH}
 \sum_H \gH{ab} \, \gstarH{a'b'} \ = \ N_f \ \delta_{aa'} \,\delta_{bb'}\ ,
\eqb
and $\Hl$ is defined by $H(l) = \{ \mu \,|\,  l_\mu \,\mbox{is odd} \}$,
$\Hh=H$ for $h$ even, $\Hh=\{\mu\,|\,\mu\not\in H\}$ for $h$ odd.
Finally \eqref{direct} becomes (summation over double spin and flavor indices
is understood)
\eqa \label{direct-result}
 \Psi_a^b(p) = \sqrt{N_f} \ \suml \ \Pi(p + \pi l) \ 
\psih_{a}^{b}(p + \pi l)\ ,\ 
 \psih_a^b(p + \pi l) = \gammh{\Hl}{aa'}\ \psi_{a'}^{b'}(p + \pi l)\
 \gdagg{\Hl}{b'b} .
\eqb
Note that the blockspin transformation is diagonal with respect to spin and
flavor for continuum momenta within the first Brillouin zone $\Bc$, 
yet not for all $l \neq 0$.

We are now prepared to compute the perfect action for a RGT of the 
Gaussian type. Starting from a continuum action with a general
mass term $m_{b}$ -- which does not need to be flavor degenerate --
the perfect lattice action $S[\bar \Psi ,\Psi ]$ is defined as
\Eqa
    &   & e^{-S[\bar \Psi ,\Psi ]}  
    \ = \ \int\!\Dpsi \int\!\Deta \ 
\ \exp \Bigl\{ -\int \!\! \frac{dq}{N_f \, \pi^d} \ \psib_a^b(-q) 
\left( i\gmu_{aa'}q_\mu+m_b \right) \psi_{a'}^b(q) \Bigr\} \nonumber\\
&\times & \exp \Bigl\{ \int_{\Bc} \frac{dp}{N_{f} \, \pi^{d}} \ \Bigl[ \ 
[ \bar \Psi _a^b(-p) \, - \, \sqrt{N_f} \suml
\Pi(p + \pi l)\ \psibh_a{}^b(-p-\pi l)\ ]\ \eta_a^b(p) \nonumber\\
& & + \etab_a^b(-p) \ [ \ \Psi_a^b(p) \, - \, 
\sqrt{N_f} \suml \Pi(p + \pi l)\ \psih_a{}^b(p + \pi l)\ ] \nonumber \\
& & + \etab_a^b(-p) D_{aa'}^{b}(p) \eta_{a'}^{b}(p) \Bigr] \Bigr\} \ ,
\label{BST}
\Eqb
where $\bar \eta$, $\eta$ are auxiliary Grassmann fields
defined on the same sites and with the same flavor structure
as $\bar \Psi$, $\Psi$.
A non-zero term 
\eqa
D_{aa'}^{b}(p) \ = \ \gmu_{aa'}D_\mu^b(p) \ + \ \delta_{aa'} \, D_0^b(p)
\eqb
``smears out'' the blockspin transformation, as in Ref.\ \cite{BBCW}. 
This term is used to optimize locality of the resulting perfect
action; it will be specified later on.
The Gaussian integrals over $\psi$,$\psib$ and $\etab$,$\eta$ can be evaluated
by substitution of the classical fields, which leads to
\footnote{We ignore constant factors in the partition function.}
\eqa \label{Sp-result}
S[\bar \Psi ,\Psi ] \ = \ \int_{\Bc} \frac{dp}{N_{f} \, \pi^{d}}
\ \Bigl[ \ \bar \Psi_a^b(-p) \
       G^{-1}{}_{aa'}^{bb'}(p) \ \Psi_{a'}^{b'}(p) \ \Bigr] \ ,
\eqb
with the lattice propagator 
\eqa \label{propagator}
    G_{aa'}^{bb'}(p) \, = \, D_{aa'}^{bb'}(p)  
 \, + \, \suml \Bigl( \Pi(p + \pi l)^2 \, 
\gammh{\Hl}{bd} \ \frac{[-i(-1)^{k_\nu}
 \gammh{\mu}{aa'} (p+ \pi l)_\mu + m_d] \, \delta_{dd'}}
 {(p+ \pi l)^2+m_d^2} \gdagg{\Hl}{d'b'} \Bigr) \ .  
\eqb
Note that $G$ is flavor diagonal, $G^{bb'}=G^{b}\delta^{bb'}$, because
$\gamma^{\Hl}\gK{}\gdagg{\Hl}{}$ is diagonal iff $\gK{}$ is. 
In particular, for a degenerate mass term the adjungation with $\gamma^{\Hl}$
is trivial, and the lattice propagator is proportional to $\delta_{bb'}$ in
flavor space. We define
\eqa \label{def-Qb} 
    G^b_{aa'}(p) \ = \ -i \sum_\mu \gmu_{aa'} \, Q^b_\mu(p)
                 \ + \ \delta_{aa'} Q^b_0(p) \ ,   
\eqb
hence $Q_\mu^b,Q_0^b$ become
\Eqa
         Q^b_\mu(p) 
   & = & D_\mu^b(p) \, + \suml \Pi(p + \pi l)^2  
 \frac{1}{N_f} \sum_{b'} \sum_K \epsilon_K(l)\, \gK{bb}\gstar{K}{b'b'}
 \frac{(-1)^{l_\mu}(p + \pi l)_\mu}
{(p + \pi l)^2+m_{b'}^2}\,,\qquad \label{Qmub}\\
         Q^b_0(p) 
   & = & D_0^b(p) \, + \suml \Pi(p + \pi l)^2 
 \frac{1}{N_f} \sum_{b'} \sum_K \epsilon_K(l)\, \gK{bb}\gstar{K}{b'b'}
 \frac{m_{b'}}{(p + \pi l)^2+m_{b'}^2}  \, . \qquad   \label{Q0b}
\Eqb 
In case of a non-degenerate mass $m_b$ the sums $\sum_{b'}\sum_K$ cannot be
contracted according to \eqref{sumH}, due to the sign factor
$\epsilon_K(l)=\epsKtext$. However,
in the degenerate case $m_b=m$, with flavor independent smearing terms
$D^{b}=D$, we simply obtain
\Eqa 
Q_\mu(p) & = & D_\mu(p) \ + \ \suml \Pi(p + \pi l)^2 \, \frac{ (-1)^{l_\mu}
(p + \pi l)_\mu}{(p+ \pi l)^2+m^2} \ ,             \label{Qmub-deg}\\
Q_0(p)   & = & D_0(p) \ + \ \suml \Pi(p + \pi l)^2  
\frac{m}{(p+ \pi l)^2+m^2} \ .               \label{Q0b-deg}
\Eqb 

The perfect action in real space arises from \eqref{Sp-result} inserting the
momentum representation given in \eqref{ytopab}. 
After some $\gamma$--matrix algebra
\cite{Dilg}, we arrive at
\Eqa
S[\bar \Psi ,\Psi ] & = & \sum_{y,y'} \bar \Psi (y) \,
{\bf m}(y,y') \, \Psi (y') \ ,           \label{SlattK1}\\
{\bf m}(y,y') & = & \sum_K \rho^K(y')\rho(y-y',y') \ 
M^K(y-y')\ . \label{SlattK2}
\Eqb
Corresponding to a lattice propagator diagonal in flavor space, the sum over
$K$ runs over multi-indices $K\in\Dc$ with diagonal $\gK{}$, in the
Weyl basis $\Dc \, = \, \{\emptyset,\ 1\,2,\ 3\,4,\ 1\,2\,3\,4\}$.
The sign factors $\rho(z,y)\equiv\rho(H(z),H(y))$ arise from 
$\gH{} \gK{} = \rho(H,K) \, \gamma^{H\Delta K}$, 
where $H\Delta K=(H\cup K)\backslash (H\cap K)$, 
and $\rho^K(y)$ is given by $\rho(H(y),K) \rho(K,H(y))$.
By symmetry, the only non-zero contributions to $M^{K}(y)$ are
\Eqa
 M^K_\mu(y) & = & i \rho(\mu,K) \int_{\Bc} \frac{dp}{\pi^{d}}
 \ e^{-ipy} \ M^K_\mu(p)
 \qquad \mbox{ for } H(y)=\mu\Delta K\ ,\qquad\label{MKy1}\\
    M^K_0(y)   & = & \int_{\Bc} \frac{dp}{\pi^{d}}
 \ e^{-ipy} \ M^K_0(p) \qquad \qquad
 \qquad \mbox{ for } H(y)=K        \ ,  \qquad\label{MKy2}\\
    \mbox{with} \quad M^K_{\mu,0}(p) 
               & = & \frac{1}{N_f} \sum_b\gstarK{bb}\,\frac{Q_{\mu,0}^b(p)}
 {\sum_\nu Q_\nu^b(p)^2 + Q_0^b(p)^2}\ .\qquad\label{def-mK}
\Eqb
The flavor degenerate case leads to vanishing components for $K\neq\emptyset$,
and (with $M^\emptyset_{\mu,0} = M_{\mu,0}$) we simply obtain
\eqa
    M_{\mu,0}(p) \ = \ \frac{Q_{\mu,0}(p)}
    {\sum_\nu Q_\nu(p)^2 + Q_0(p)^2} \ . \label{def-mK-deg}
\eqb

It has been proven in Ref.\ \cite{Dilg} that the couplings given by the fermion
matrix ${\bf m}(y,y')$ are local, i.e.\ they decay faster than any power of
$|y-y'|$. 
For that, certain periodicity properties apply, which translate into (for
simplicity of notion, $\mut$ denotes either $\mu$ or $0$, and
$K\Delta0\equiv K$)
\eqa \label{periodicity}
    Q^b_{\mut}(p) \ = \ \sum_{K\in\Dc} \gK{bb} \ Q^K_{\mut}(p) \ , \quad
    Q^K_{\mut}(p+ \pi \hat{\nu}) \ = \ 
    \twocase{ -Q^K_{\mut}(p) }{ \nu \,     \in \ K\Delta\mut }
            {  Q^K_{\mut}(p) }{ \nu \, \not\in \ K\Delta\mut } \ .
\eqb
Again, we sum over diagonal $\gamma$--matrices only. It is provided that
the corresponding requirements are met for the smearing terms $D^b_{\mut(p)}$
within $Q^b_{\mut}(p)$, see below. 
In consequence, the fermion matrix components $M^K_{\mut}(p)$ obey periodicity
conditions analogous to $Q^K_{\mut}(p)$, and the integrands of 
\eqRef{MKy1}{MKy2}
are periodic with respect to the Brillouin zone $\Bc$ and analytic in a
strip around the real axis. This implies locality of the perfect action.

The coupling of even and odd lattice points is due to 
the $M^K_\mu$ components of
the fermion matrix; the $M^K_0$ components couple even--even and odd--odd. 
We add without proof that even-odd decoupling of the Hermitian matrix
${\bf m}^\dagger\, {\bf m}$ 
can be shown in any even dimension $d$ with arbitrary
(non-degenerate) mass terms for truncated versions of the
perfect fermion matrix ${\bf m}(y,y')$, see Ref.\ \cite{Dilg} for $d=2$. 
This is a useful property in simulations
with Hybrid Monte Carlo algorithms. However, with (non-perfect) coupling to a
gauge field and non-zero mass term (i.e.\ with even--even, odd--odd as well as
even--odd couplings), this is not true in general. 
\footnote{Since even--odd decoupling of 
${\bf m}^\dagger\,{\bf m}$ is a perfect property, it could be 
imposed as a construction requirement for an approximately
perfect  fermion-gauge vertex. For $m=0$, even--odd decoupling
of ${\bf m}^\dagger\,{\bf m}$ is guaranteed (${\bf m}$
only couples even with odd sites).}

%\clearpage
%%%%%%%%%%%%%%%%%%%%%%%%%%%%%%%%%%%%%%%%%%%%%%%%%%%%%%%%%%%%%%%%%%%%%%%%%%%
\section{Optimization of locality} \label{Locality}

In the blocking scheme described so far, there is quite some freedom
left. In particular, 
we may use averaging functions different from 
%the standard block average given by 
$\Pi_{BA}$, % in \secref{Scheme}
and we can choose the smearing term $D$ in \eqref{BST}. 
In both cases we aim at optimization of the locality in the resulting 
perfect action. 

Let us first discuss the averaging scheme. In \eqref{BST} we implicitly assumed
the same blocking of $\psi$ and $\psib$, given by the weight function
$\Pi(x)$ resp.\ its Fourier transform $\Pi(p)$. Now we consider the case of a
block average $\Pi=\Pi_{BA}$ for $\psi$ ($\psib$) only, while $\psib$ 
($\psi$) is put on the lattice by decimation, $\Pi(x)=\delta
(x)$. Thus we obtain a single factor 
$\Pi(p + \pi l)$ in \eqref{propagator} with
\eqa
 \Pi(p)  \ = \ \Pi_{BA}(p) \ = \ \prod_\mu \frac{\ph_\mu}{p_\mu} \ , \qquad
 \ph_\mu \ = \ \sin p_{\mu} \ .
\eqb
In case of a $\delta$ function RGT ($D=0$), this 
means to identify the averaged continuum and lattice 2-point functions
\eqa \label{twopoint}
 \langle \ \phi(y) \ \phib(y') \ \rangle \ = \ 
 \int_{[y]}dx\ \langle \ \varphi(x,H(y)) \ \varphib(y',H(y')) \ \rangle \ .
\eqb
Due to translation invariance it does not matter whether we average source or
sink of the continuum expression, or whether we allocate the space directions
to be integrated over to source and sink in some way. (The last point of view 
may be used to make a closer contact to the construction of staggered fermions
from DK fermions in the continuum \cite{BJ}, as discussed in 
Ref.\ \cite{Dilg}.)
We call this blocking scheme {\em partial decimation}. For
the 2-point functions every space direction is integrated over once; 
therefore we do not run into the difficulties arising for
blockspin transformations with 
complete decimation, which do not have a corresponding perfect action.
%which would not even be well-defined for blocking from the
%continuum, i.e.\ with infinite blocking factor.

For both blocking schemes, block average for $\psi$ and $\psib$ (BA) and
partial decimation (PD), we now want to optimize locality of the
couplings by making use of the smearing terms in \eqRef{Qmub}{Q0b}.
As an optimization criterion it has been suggested to require
that in the effectively 1d case -- with momenta
$p=(p_{1},0,\dots ,0)$ -- the couplings are restricted to nearest
neighbors as in the standard action \cite{QuaGlu,Dallas}.%
\footnote{It has been shown in Ref.\ \cite{scal} that this 
criterion does optimize locality in $d=4$ for scalar fields
over a wide range of masses.}
In the degenerate case we require
\eqa
   i\gamma_1 \, M_{1}(p_{1},0,\dots ,0) \ + \ M_{0}(p_{1},0,\dots ,0) \ = \  
   f(m) \, [ \, i \hat p_{1} \, \gamma_{1} \ + \ \hat m \, ] , \label{requ}
\eqb
with $\hat m \vert_{m=0}=0$ and $f(0)=1$.
Our ansatz for the Gaussian smearing term reads
\begin{equation}
D_{\mu}(p) = c(m)\hat p_{\mu} , \qquad D_{0}(p) = a(m).
\end{equation}
Requirement (\ref{requ}) can be fulfilled in both blocking schemes
we are considering, if we specify the RGT as follows
\begin{eqnarray}
c_{BA}(m) & = & [\cosh m -1] \, / \, (2m)^2 \ , \quad
a_{BA}(m) \ = \ [\sinh(2m) -2m]    \, / \, (2m)^2 \ , \nonumber\\ 
c_{PD}(m) & = & 0 \ , \qquad \qquad \qquad \qquad \qquad \!
a_{PD}(m) \ = \ [\cosh m -1] \, / \, (2m)   \ . \label{locop}
\end{eqnarray}
In both cases, we obtain $\hat m = \sinh m $.
For $m=0$ a non-vanishing static smearing term $a(0)$
would explicitly break the remnant chiral symmetry in the
fixed point action. Therefore, the static term should vanish 
for optimized locality, as it does in both cases.
Furthermore, in the PD scheme the chiral limit is optimized 
for locality by
a simple $\delta$ function RGT. As an advantage of this property --
which is not provided by the BA scheme -- there is a direct relation
between the $n$-point functions in the continuum and on the
lattice. In addition, the extension to interacting theories
might involve numerical RGT steps in the classical limit, 
which also simplify in the absence of a Gaussian smearing term.
Finally, a non-vanishing term $c(m)$ causes complications
if one wants to include a gauge interaction in the RGT, 
but the PD scheme avoids such problems.

The decay of the couplings ${\bf m}(x,0) = {\bf m}(x)$
in the massless case for $d=2$ and $d=4$ is
shown in Figures 1 and 2.
%\begin{figure}[htb]
%\begin{center}
%\epsfig{file=coupl0-0-2d.ps,width=0.5\textwidth}
%\epsfig{file=coupl0-0-4d.ps,width=0.5\textwidth}
%\parbox{\textwidth}{ \caption{\label{coupl00} \sl 
%The decay of couplings in (1,0) direction for $d=2$ (left) and $d=4$ (right) 
%in the massless case: BA without smearing (diamonds), BA optimized (squares), 
%PD (triangles).
%}}
%\end{center}
%\end{figure}
%figalt - Ende
%%%%%%%%%%%%%%%%%%%%%%%%%%%%%%%%%%%%%%%%%%%%%%%%%%%%%%%%%%%%%%%%%%%%
% - included  ps-figure -------------------------
\begin{figure}[htb]
\begin{center}
\epsfig{file=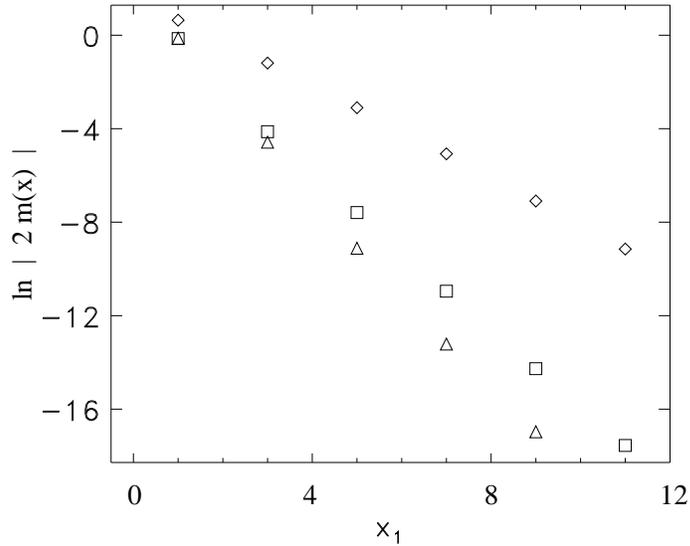,width=\figwidth}
\parbox{\textwidth}{ \caption{\label{coupl00-2d} \sl 
The decay of couplings in (1,0) direction for $d=2$ in the massless case: 
BA without smearing (diamonds), BA optimized (squares), PD (triangles).
}}
\end{center}
\end{figure}
%%%%%%%%%%%%%%%%%%%%%%%%%%%%%%%%%%%%%%%%%%%%%%%%%%%%%%%%%%%%%%%%%%%%
% - included  ps-figure -------------------------
\begin{figure}[htb]
\begin{center}
\epsfig{file=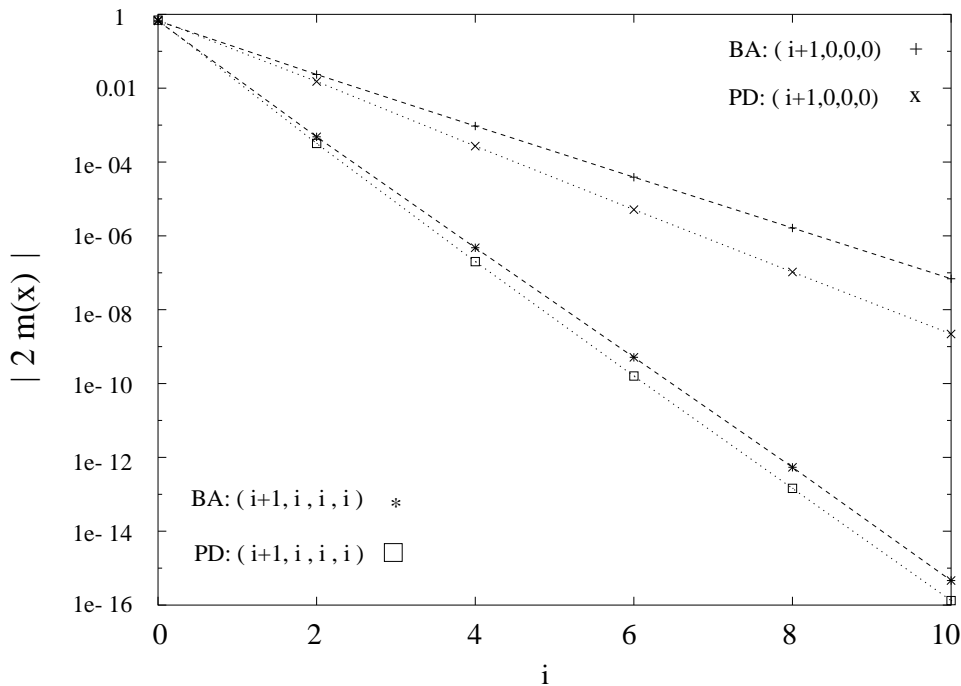,width=\figwidth,angle=270}
\parbox{\textwidth}{ \caption{\label{coupl00-4d} \sl 
The decay of 4d couplings in various directions in the massless case: 
BA with optimal smearing term and PD. We see that the latter
couplings decay faster.}}
\end{center}
\end{figure}
%%%%%%%%%%%%%%%%%%%%%%%%%%%%%%%%%%%%%%%%%%%%%%%%%%%%%%%%%%%%%%%%%%%%
We see that the PD blocking scheme works better. In the 
non-degenerate case, the simplest ansatz for a smearing term is 
(corresponding to
the staggered fermion action with non-degenerate mass \cite{non-deg})
\eqa
 D^b_\mu(p) \ = \ c \, \ph_\mu \ , \qquad 
 D^b_0(p)\ =\ \sum_{K\in\Dc} \gK{bb}\ a_K \prod_{\mu\in K} \cos p_\mu \ .
\eqb
In this case, we obtained -- by means of numerical optimization -- a similar 
decay of couplings as with degenerate masses, yet no strict 1-dimensional
ultralocality. For $d=2$ it appears that a non-degenerate parameter 
$a_{12} \neq 0$ does not improve the coupling decay significantly, so we
worked with $a_{12}=0$. Again the PD scheme, where we assumed $c_{PD}=0$, 
leads to a more local action, see Figure 3.
\begin{figure}[htb]
\begin{center}
\epsfig{file=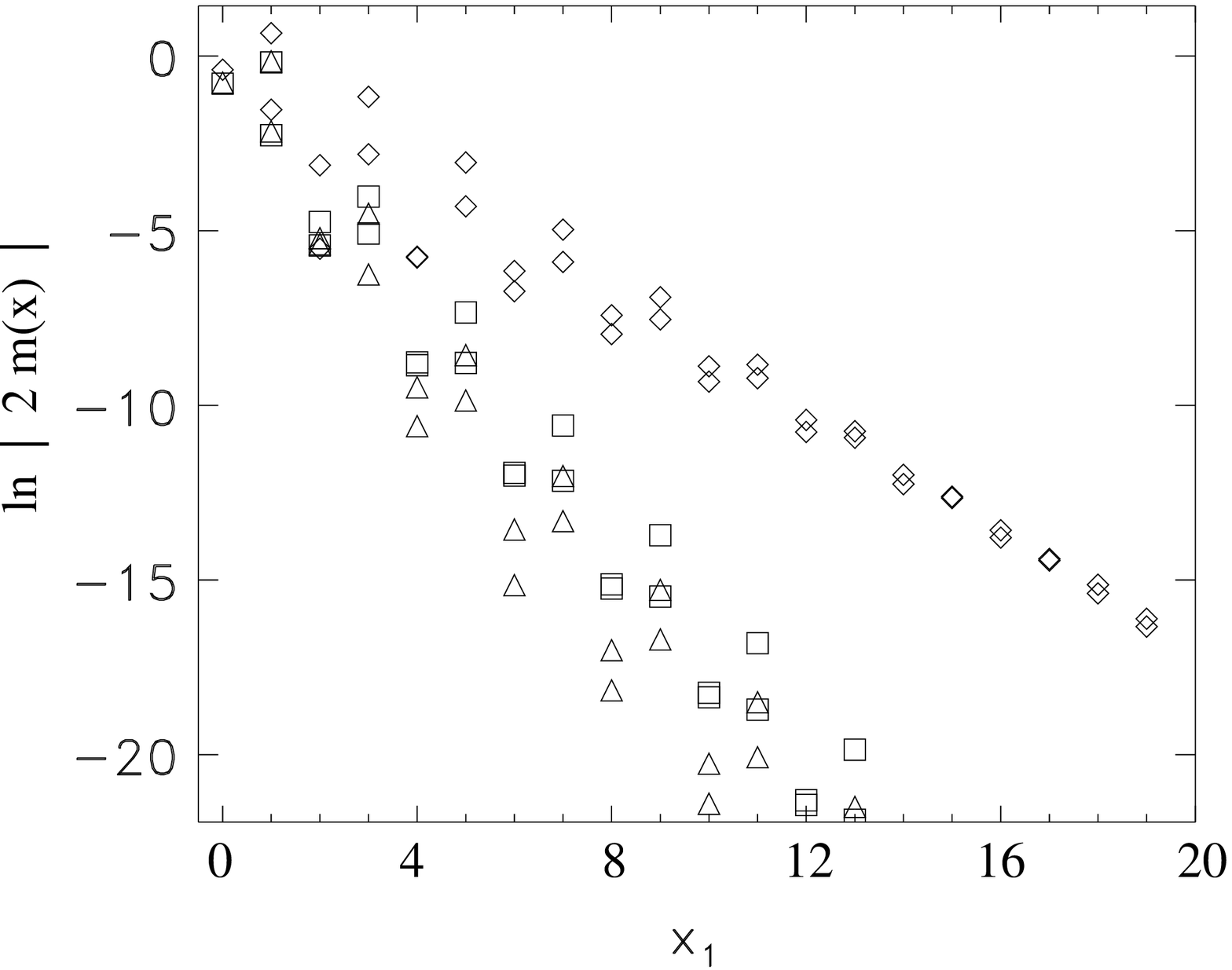,width=\figwidth}
\parbox{\textwidth}{ \caption{\label{coupl01-2d} \sl 
The decay of couplings in (1,0) direction for $d=2$ with masses
$m_1 = 0$, $m_2 = 0.5$: BA without smearing (diamonds), 
BA optimized (squares), PD (triangles).
}}
\end{center}
\end{figure}

%\clearpage
%%%%%%%%%%%%%%%%%%%%%%%%%%%%%%%%%%%%%%%%%%%%%%%%%%%%%%%%%%%%%%%%%%%%%%%%%%%
\section{Truncation effects} \label{Truncation}

The litmus test of any perfect action is given by the truncation effects in
a practicable number of remaining couplings. These effects are minimized by
maximal locality. Yet, the truncation scheme itself may have some impact on
the truncation errors. An elegant procedure has been proposed in 
Ref.\ \cite{LAT96} for Wilson fermions.
First the perfect action is constructed  on a small lattice
volume $N^{d}$ by restricting the momentum components to the discrete
values $p_{\mu} = 2\pi n_{\mu}/N \in ]-\pi ,\pi ]$, $n_{\mu} \in \Z$.
Typically, one chooses $N=3$, and the resulting couplings 
are then used in a large volume too, 
where they are not exactly perfect any more.
This truncation scheme has the virtues of automatically correct
normalizations, and a simplification of the 
numerical evaluation as opposed to a truncation in coordinate space.
Furthermore, the mapping to the corresponding truncated perfect
action in a lower dimension remains exact; for instance, one
can reproduce the effectively 1d nearest neighbor action starting
from $d>1$ by summing over the extra dimensions, which provides
a sensitive test of the numerical accuracy.
Provided a good locality, this method works well for Wilson-type
fermions, gauge fields \cite{LAT96} and scalars \cite{scal}.

For staggered fermions, only a crude truncation in coordinate 
space has been applied so far \cite{BBCW}. 
One includes the coupling distances $\pm 1,\ \pm 3$ in the $\mu$ direction,
and $2,0,-2$ in the non-$\mu$ directions. This involves
more couplings than the $N=3$ truncated Wilson-type fermion
(called ``hypercube fermion''), whereas the number of degrees of freedom
is the same in both cases.
% hd- Nevertheless, the quality of its spectral
% and thermodynamic properties did not reach the level of the
% ``hypercube fermion'', although it was improved over the standard
% staggered formulation. Here we try to achieve such a high
% quality also for a truncated perfect staggered fermion,
% based on increased locality and also on an improved 
% truncation scheme, closer to the elegant procedure described above.
However, although significant improvement has been achieved compared to the
standard staggered formulation, the quality of its spectral
and thermodynamic properties did not reach the level of the
``hypercube fermion''. 
In order to arrive at results of a similar quality, 
we adapt the above truncation 
procedure -- together with the PD scheme -- to the case of staggered fermions.

\begin{figure}[htb]
\begin{center}
\epsfig{file=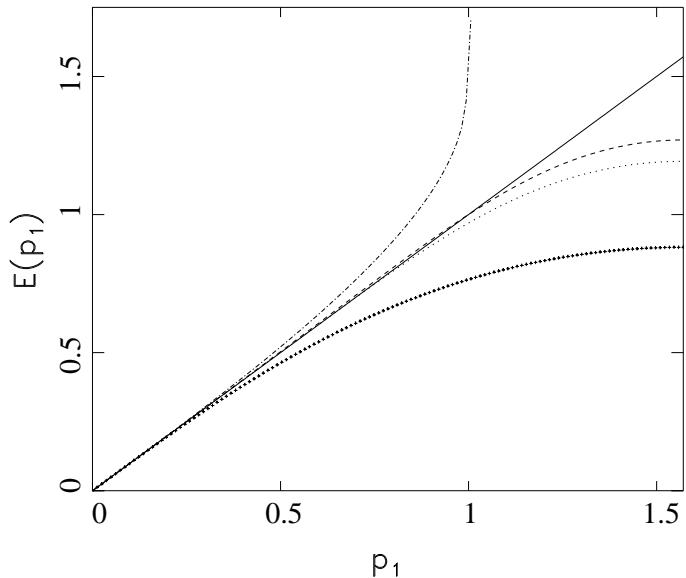,width=\figwidth}
\parbox{\textwidth}{ \caption{\label{spectrum2d-00} \sl 
The massless spectrum in $d=2$. 
We compare standard staggered fermions (crosses) and 
perfect spectrum (full line) with perfect actions truncated  with $N=3$ 
and mixed periodic boundary conditions: block average 
(dotted), partial decimation (dashed). The last case is also 
plotted for truncation with periodic boundary conditions (dashed-dotted).
}}
\end{center}
\end{figure}

We treat the components of the fermion matrix $M^K_{\mut}(p),\mut=\mu,0$
separately, as they show different behavior under reflections, 
%$\Pi^\nu: y_\nu,p_\nu \rightarrow -y_\nu,-p_\nu$
\eqa \label{reflections}
M^K_{\mut}(p_{1}\dots p_{\nu -1},-p_{\nu},p_{\nu+1} \dots p_{d}) \ = \ 
    \twocase{ -M^K_{\mut}(p) }{ \nu \,     \in \ K\Delta\mut }
            {  M^K_{\mut}(p) }{ \nu \, \not\in \ K\Delta\mut } \ .
\eqb 
Again we use the short-hand notation $K\Delta\mut=K$ for $\mut=0$. 
Truncation is achieved by discrete Fourier transformation and a discrete
support given by $c_N(y_\mu)=1,1/2,0$ for $| y_\mu|<,=,>N$, respectively,
\eqa
 M^K_{\mut;N}(y) \ = \ \prod_\mu \frac{\pi c_N(y_\mu)}{N} \
 \sum_{p \in \Bc^K_{\mut;N}} e^{-ipy} M^K_{\mut}(p) \ .
\eqb
It is the set of discrete momenta $\Bc^K_{\mut;N}$ which depends on the
reflection properties given by $\mut,K$. We choose
mixed periodic and antiperiodic boundary conditions,
\eqa
 p \in \Bc^K_{\mut;N} \ \Leftrightarrow \ 
 p_\nu = \twocase{(n+1/2)\pi/N}{\nu \,     \in \ K\Delta\mut}
       {n\pi /N}{\nu \, \not\in \ K\Delta\mut}, \ n\in\Z,\ \
 p_\nu \in \ ]-\pi /2,\pi /2] \ .
\eqb
It is easily verified that the transformation of the truncated components 
back to momentum space reproduces the perfect values at 
$p\in\Bc^K_{\mut;N}$ ,
\eqa
 M^K_{\mut;N}(p) \equiv \sum_y e^{ipy}\,M^K_{\mut;N}(y) \ = \  M^K_{\mut}(p)
 \quad \mbox{ for } \ p \in \Bc^K_{\mut;N} \ . 
\eqb 
The components $M^K_{\mut}(y)$ inherit the reflection behavior of
$M^K_{\mut}(p)$ described in \eqref{reflections}.
Therefore, treating them as
periodic functions in $y$--space by discrete momenta $p_\nu=\pi n/N,n\in\Z$ 
would make them vanish artificially on the boundaries $y_\nu = \pm N$ for 
$\nu\in K\Delta\mut$. We avoid this defect by the
above choice of discrete momenta. It pays off
by a drastic reduction of truncation 
effects, see \fig{spectrum2d-00} for the 2d spectrum in the 
massless case using partial decimation. 
%\begin{figure}[htb]
%\begin{center}
%\epsfig{file=spec0-0ext.ps,width=0.5\textwidth}
%\epsfig{file=spec0-1.ps,width=0.5\textwidth}
%\parbox{16cm}{ \caption{\label{spectrum2d} \sl 
%The $d=2$ spectrum for $m_{1,2}=0$ and $m_1=0, m_2=1$. We compare standard 
%staggered (crosses) and perfect spectrum (full line) with effective actions 
%truncated  with $N=3$ and partially antisymmetric boundary conditions: block 
%average (dashed--dotted), partial decimation (dashed). The last case is also
%plotted for truncation with periodic boundary conditions (dotted).
%}}
%\end{center}
%\end{figure}
%figalt - Ende
\begin{figure}[htb]
\begin{center}
\epsfig{file=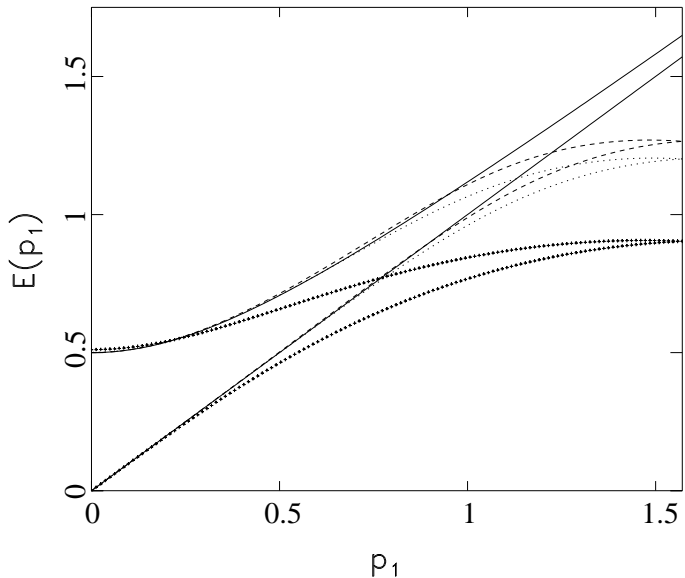,width=\figwidth}
\parbox{\textwidth}{ \caption{\label{spectrum2d-01} \sl 
The spectrum in $d=2$ for $m_1=0, m_2=0.5$. 
Standard staggered (crosses) and perfect spectrum (full line) 
are compared with perfect actions truncated  with $N=3$
mixed periodic boundary conditions: block average (dotted), 
partial decimation (dashed). 
}}
\end{center}
\end{figure}
%%%%%%%%%%%%%%%%%%%%%%%%%%%%%%%%%%%%%%%%%%%%%%%%%%%%%%%%%%%%%%%%%%%%
The values of the perfect couplings, truncated by {\em mixed
periodic boundary conditions} (as described above) for $N=3$, 
are given in Table \ref{tabcop}.
\begin{table}
\begin{center}
\begin{tabular}{|c|c|c|c|c|c|}
\hline
$d=4$ & & & $d=2$ & & \\
\hline
& PD & BA & & PD & BA \\
\hline
\hline
(1,0,0,0) & {\bf 0.348194} &  0.331558 & (1,0) & {\bf 0.4391168527} 
&  0.433150 \\
\hline
(1,2,0,0) & {\bf 0.020490} &  0.022963 & (1,2) & {\bf 0.0304416413}
&  0.033425 \\
\hline
(1,2,2,0) & {\bf 0.002240} &  0.002435 & & & \\
\hline
(1,2,2,2) & {\bf 0.000247} &  0.000180 & & & \\
\hline
(3,0,0,0) & {\bf 0.007609} &  0.011721 & (3,0) & {\bf 0.0052123698}
&  0.008022 \\
\hline
(3,2,0,0) &{\bf -0.000216} & -0.000319 & (3,2) &{\bf -0.0026062073}
& -0.004011 \\
\hline
(3,2,2,0) &{\bf -0.000384} & -0.000605 & & & \\
\hline
(3,2,2,2) &{\bf -0.000214} & -0.000317 & & & \\
\hline
\hline
\end{tabular}
\end{center}
\caption{\sl The couplings of the perfect action for massless staggered
fermions, constructed from the ``partial decimation'' (most successful)
and from the optimized ``block average'' scheme,
and truncated by mixed periodic boundary conditions.
The couplings are odd in the odd
component, and even in all other components. Among the latter
there is also permutation symmetry.
The 2d PD couplings (given to higher accuracy)
are used in the Schwinger model simulation reported below.}
\label{tabcop}
\end{table}

Furthermore, the truncation effects are significantly stronger for 
even truncation distances $N$; in particular, for $N=2$ and $N=4$
the energies become complex-valued at large momenta.
 
For $N=3$ -- which is still tractable in numerical
simulations -- the
amount of improvement is already striking. We compare the spectra for
$d$\,=\,2 in the PD and BA blocking scheme in the massless case
(\fig{spectrum2d-00}) and 
for non-degenerate masses $m_1=0, m_2=0.5$ (\fig{spectrum2d-01}).
The couplings derived from the PD blocking scheme turn out to be better,
in agreement with the higher degree of locality observed in the (untruncated)
perfect action.
%, see Fig.~(\ref{coupl00-2d},\ref{coupl01-2d}).
We see that the improvement is still good in the case of non-degenerate 
masses. In this case, we optimized the smearing parameters numerically,
as pointed out in \secref{Locality}. The values are $a_{BA}=0.120$, 
$c_{BA}=0.125$, $a_{PD}=0.082$, $c_{PD}=0$.
The standard staggered fermion results for non-degenerate 
masses are calculated with the action proposed in Ref.\ \cite{non-deg}.
We emphasize that the spectra of (untruncated) perfect actions
are indeed perfect, i.e.\ identical to the continuum spectra
(up to periodicity, which is inevitable on the lattice).
Hence the spectrum reveals directly the artifacts due to truncation.

%%%%%%%%%%%%%%%%%%%%%%%%%%%%%%%%%%%%%%%%%%%%%%%%%%%%%%%%%%%%%%%%%%%%%%%%% 
Figure 6 compares the massless spectra in $d=4$, 
and we see that the qualitative behavior observed in $d=2$ persists.
We compare the standard staggered fermion, 
and the optimized BA and PD fixed point fermions, 
both truncated by $N=3$ mixed periodic boundary conditions.
We see again that the PD scheme is superior.
Furthermore, we show for comparison also the dispersion relation
of a Symanzik improved action called ``p6'' from Ref.\ \cite{Bielef}. 
Symanzik improvement by additional couplings along the axes
(Naik fermion, \cite{Naik}) only yields a moderate quality
\cite{BBCW}, but the Bielefeld group suggests a number of
actions, where Symanzik improvement is achieved by diagonal
couplings. The p6 action is the best variant among them,
and also Figure 6 confirms its excellent level of improvement.
However, it is not obvious how that formulation can
include a general mass term in a subtle way.
\begin{figure}[htb]
\begin{center}
\epsfig{file=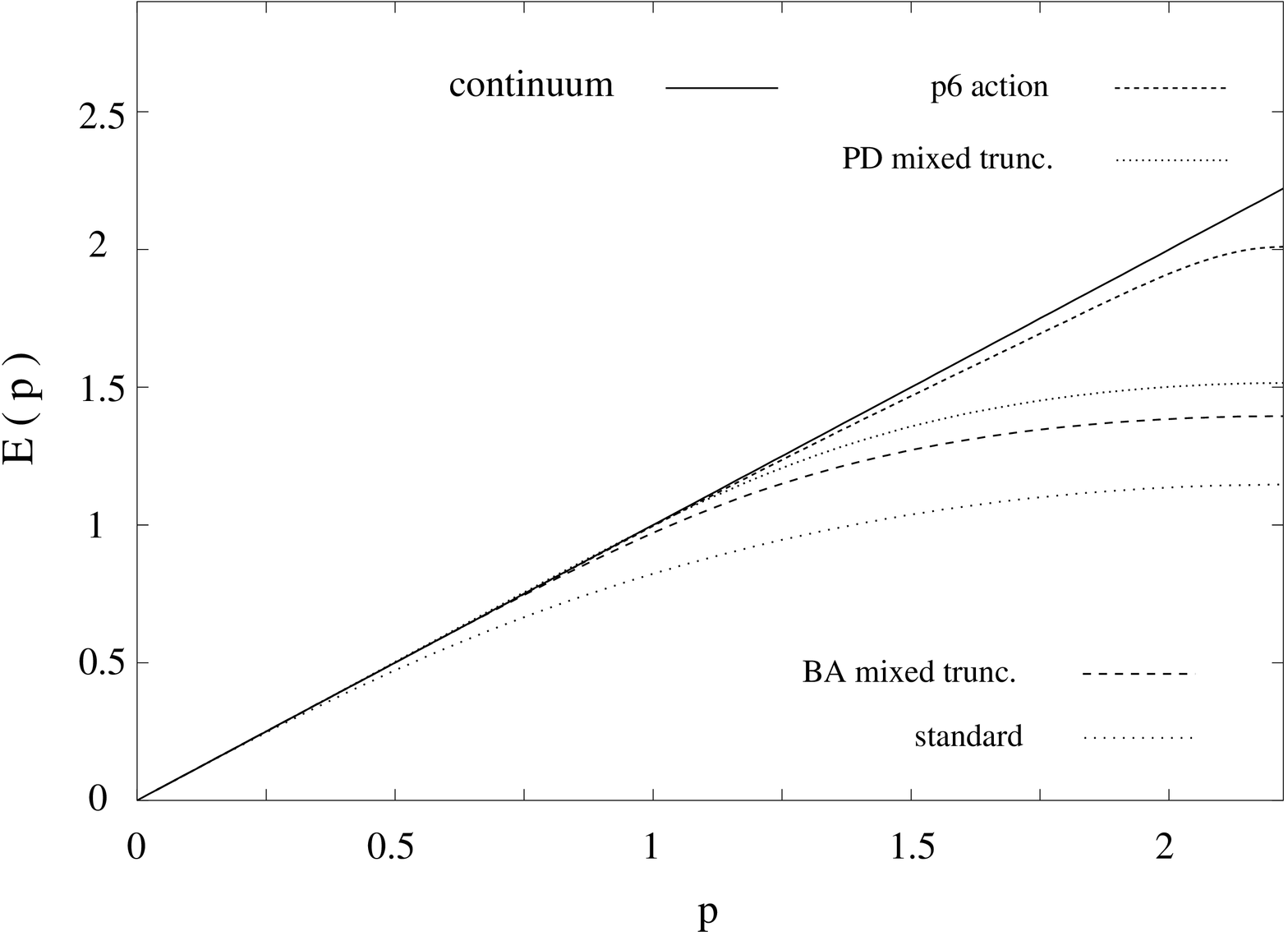,width=\figwidth,angle=270}
\end{center}
\parbox{\textwidth}
{ \caption{\sl The dispersion relation for various types of 
4d massless staggered
fermions along (1,1,0). We compare the standard action, 
a Symanzik improved action called ``p6'' from Ref.\ [8] %\cite{Bielef},
and the optimized BA and PD fixed point actions, truncated
with mixed periodic boundary conditions.}}
\label{spec4d}
\end{figure}

As a further test in $d=4$, we consider two thermodynamic
scaling quantities. The first is the
ratio $P/T^{4}$ ($P$: pressure, $T$: temperature). According
to the Stefan-Boltzmann law, this ratio is $7\pi^{2} /180$ 
for massless fermions in the
continuum, and a lattice action with many discrete points $N_{t}$
in the temporal direction will asymptotically reproduce that
value. However, the speed of convergence, and in particular
the behavior at small $N_{t}$, depend on the quality
of the action. In contrast to the spectrum, this ratio is not even 
exact for the fixed point action in \eqRef{Qmub-deg}{Q0b-deg}, because
of the ``constant factors'' that we ignored when performing the
functional integral in the RGT \eqref{BST}. 
Such factors may depend on the
temperature, so with respect to thermodynamics our action is not
fully renormalized \cite{LAT96}. However, it turns out that
the unknown factor is very close to 1, except for the regime
of small $N_{t}$ (about $N_{t}\leq 6$), which corresponds to
very high temperature. So the main issue is again the 
contamination due to the truncation.

In Figure 7 we compare this thermodynamic scaling
for a variety of staggered fermion actions at $m=0$,
and again the PD scheme turns out to be very successful.
A similar level of improvement can be observed for the p6 action,
which is also here by far better than the Naik fermion. We conclude that
a good improvement method should in any case include diagonal
couplings, which are far more promising than additional couplings on
the axes (just consider rotational invariance, for example).
%Thus we can reduce the value of $N_{t}$, which is needed to see
%scaling setting in, 
Thus we can expand the range, in which a practically
accurate continuum behavior is observed,
by about a factor of 4 compared to standard staggered fermions
(see Figures 4, 5, 6 and 7).
%including also some
%improvements \`{a} la Symanzik, but the truncated perfect action
%in the PD scheme is again most successful. The quality of a Symanzik
%staggered fermion action, which is Symanzik improved by additional
%couplings on the axes (Naik fermion \cite{Naik}) is not rather
%moderate \cite{BBCW}, but an excellent level of improvement can also
%be achieved by Symanzik improvement involving diagonal couplings;
%as an example, we also show in Figure \ref{presfig} the so-called ``p6''
%action from Ref.\ \cite{Bielef}.
\begin{figure}[htb]
\begin{center}
\epsfig{file=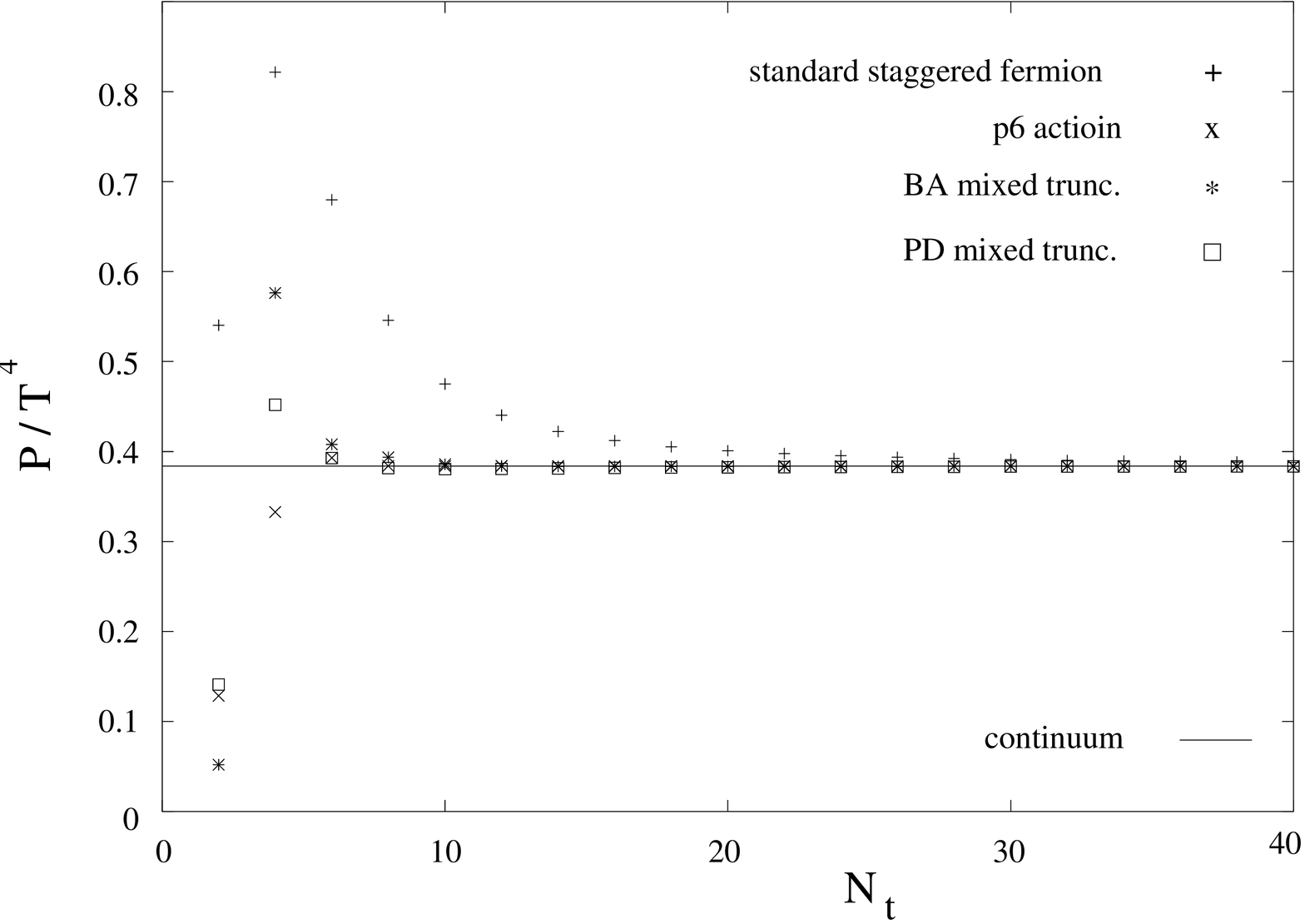,width=\figwidth,angle=270}
\parbox{\textwidth}{ \caption{\sl 
The Stefan-Boltzmann law for standard staggered fermions
and for the ``p6 action'', compared to
the truncated perfect fermions for the optimized BA and PD scheme.}}
\label{presfig}
\end{center}
\end{figure}

As a second example for thermodynamic scaling of free
massless fermions, we set $T=0$ but switch on a chemical potential
$\mu$. The ratio of the ``baryon number density'' $n_{B}$
divided by $\mu^{3}$ is our second scaling quantity.
If $\mu$ approaches 0, this ratio converges to its
continuum value $2/(9 \pi^{2})$ for any lattice action.
For increasing $\mu$ the lattice artifacts become visible,
and we see again that the truncated perfect staggered
fermions, especially those based on the PD scheme,
stay much longer in the vicinity of the continuum value,
see Figure 8. Again the practically accurate regime is
extended by a factor~\raisebox{-.4ex}{$\stackrel{>}{\sim}$}~4.
\begin{figure}[htb]
\begin{center}
\epsfig{file=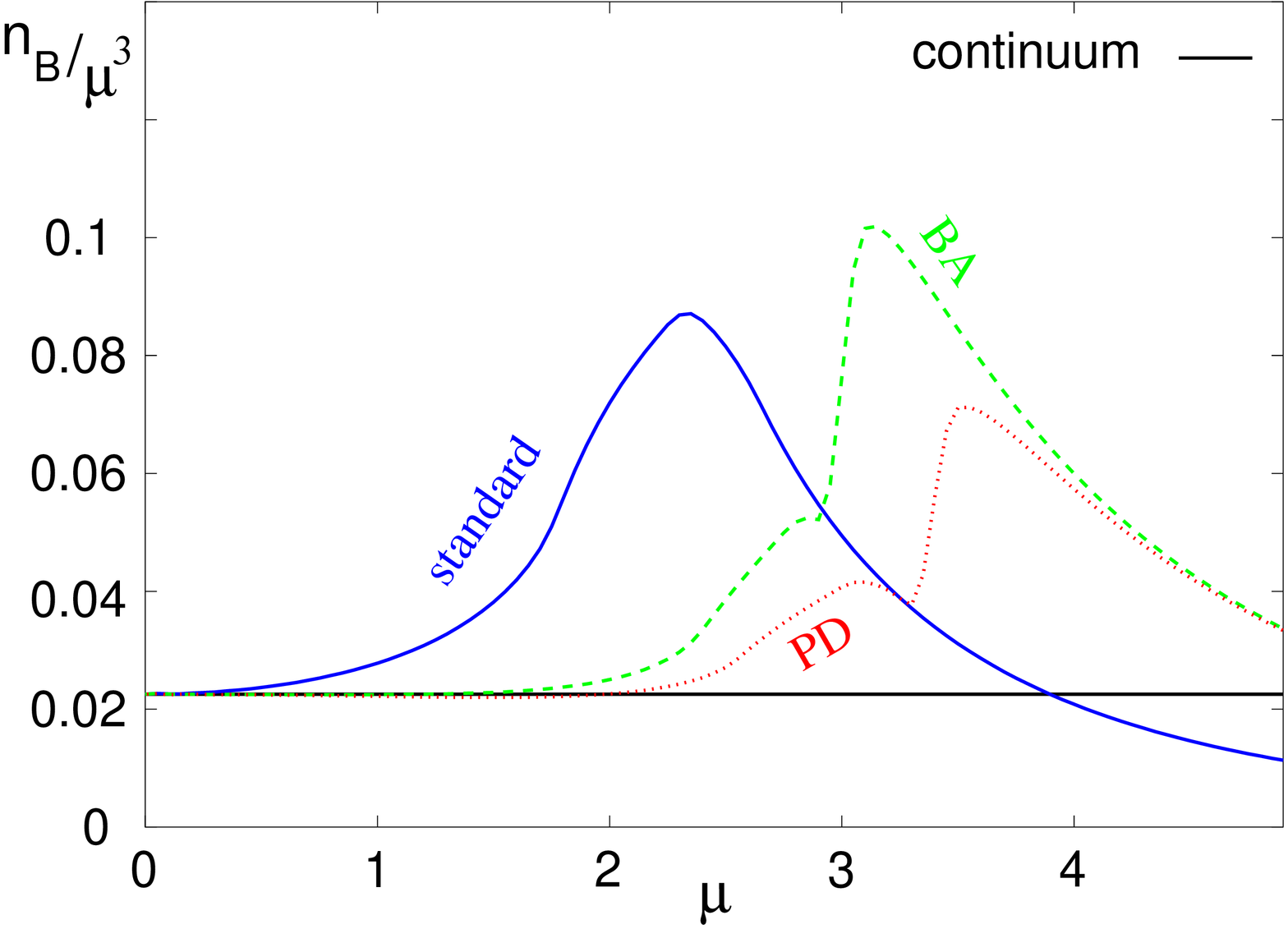,width=\figwidth,angle=270}
\parbox{\textwidth}{ \caption{\sl 
The scaling ratio $n_{B}/\mu^{3}$ at $T=0$ for different
types of free staggered fermions, as a function of the chemical
potential $\mu$.}}
\label{baryfig}
\end{center}
\end{figure}

%\clearpage
\section{The fermion-gauge vertex function} \label{Model}

\begin{figure}[htb]
\begin{center}
\vspace{5mm}
\input{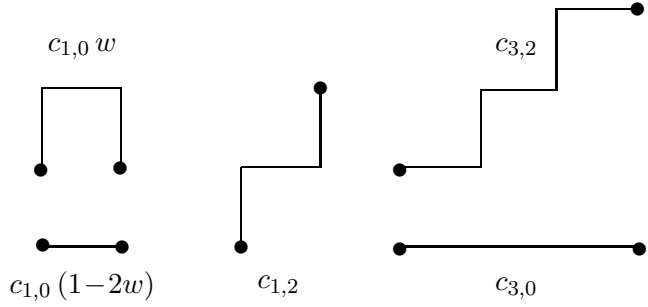}

\vspace{5mm}

\parbox{\textwidth}{ \caption{\label{coupl-fig} \sl 
The parallel transporters representing the $5$ independent couplings. 
}}
\end{center}
\end{figure}

We construct a quasi-perfect fermion-gauge vertex using the truncated
perfect couplings of the free fermions based on the PD scheme,
given in Table 1.
Without a mass term, only even--odd couplings are present. 
The coupling to the gauge field is 
achieved by the insertion of $U(1)$ parallel 
transporters on shortest lattice paths between the sites of $\bar \Psi$ 
and $\Psi$. 
Where several shortest ways exist, we use only staircase-like paths, as
illustrated in \fig{coupl-fig}.
For $\bar \Psi$ and $\Psi$ on nearest neighbor sites we also allow
for the shortest detour (staple) with some weight factor $w$, 
leading to ``fat links''.
%The corresponding staple weight $w$ is optimized effectively, see below.
Therefore, the fermionic part of the action takes the form
\Eqa
    S_f[\bar \Psi, \Psi ,U] 
    & = & \sum_{x} \sum_\mu \rho(\mu,x) \ \Bigl\{ \ 
           c_{1,0} \ (1-2w)\ \bar \Psi (x) \ U_\mu(x) \ 
\Psi (x+e_\mu) \nonumber\\
    & + & c_{1,0} \ w \sum_{\nu} \bar \Psi (x)\ 
U_\nu(x)U_\mu(x+e_\nu) 
                   U_{-\nu}(x+e_\nu+e_\mu) \ \Psi (x+e_\mu) \nonumber\\
    & + & c_{1,2} \sum_\nu \bar \Psi (x) \ U_\nu(x) U_\mu(x+e_\nu) 
                  U_\nu(x+e_\nu+e_\mu) \ \Psi (x+e_\mu+2e_\nu) \nonumber\\
\Stm& + & c_{3,0} \ \bar \Psi (x) \ U_\mu(x) U_\mu(x+e_\mu) U_\mu(x+2e_\mu) \ 
                  \Psi (x+3e_\mu) \nonumber\\
\Str& + & c_{3,2} \sum_\nu \bar \Psi (x) \ U_\mu(x) U_\nu(x+e_\mu) 
                  U_\mu(x+e_\mu+e_\nu) U_\nu(x+2e_\mu+e_\nu) \nonumber\\
    &   &         \qquad\times\qquad\Bigl. U_\mu(x+2e_\mu+2e_\nu) \ 
                  \Psi (x+3e_\mu+2e_\nu) \ \Bigr\}  \ ,      \label{Sf} 
\Eqb
where we sum over $\mu=\pm1,\pm2$, $\nu=\pm\mub$, 
and $\mub$ is the direction perpendicular to $\mu$. 
For positive $\mu$ the sign factor reads
\eqa
    \rho(\mu,x) \ = \ (-1)^n, \quad n \ = \ \sum_{\nu < \mu} x_\nu \ ,
\eqb
and negative $\mu$ follow from $e_{-\mu}=-e_\mu$, 
$\rho(-\mu,x)=-\rho(\mu,x)$, $U_{-\mu}(x)=U_\mu(x-e_\mu)^*$.

\begin{figure}[htb]
%\hspace{2cm}
\begin{center}
\epsfig{file=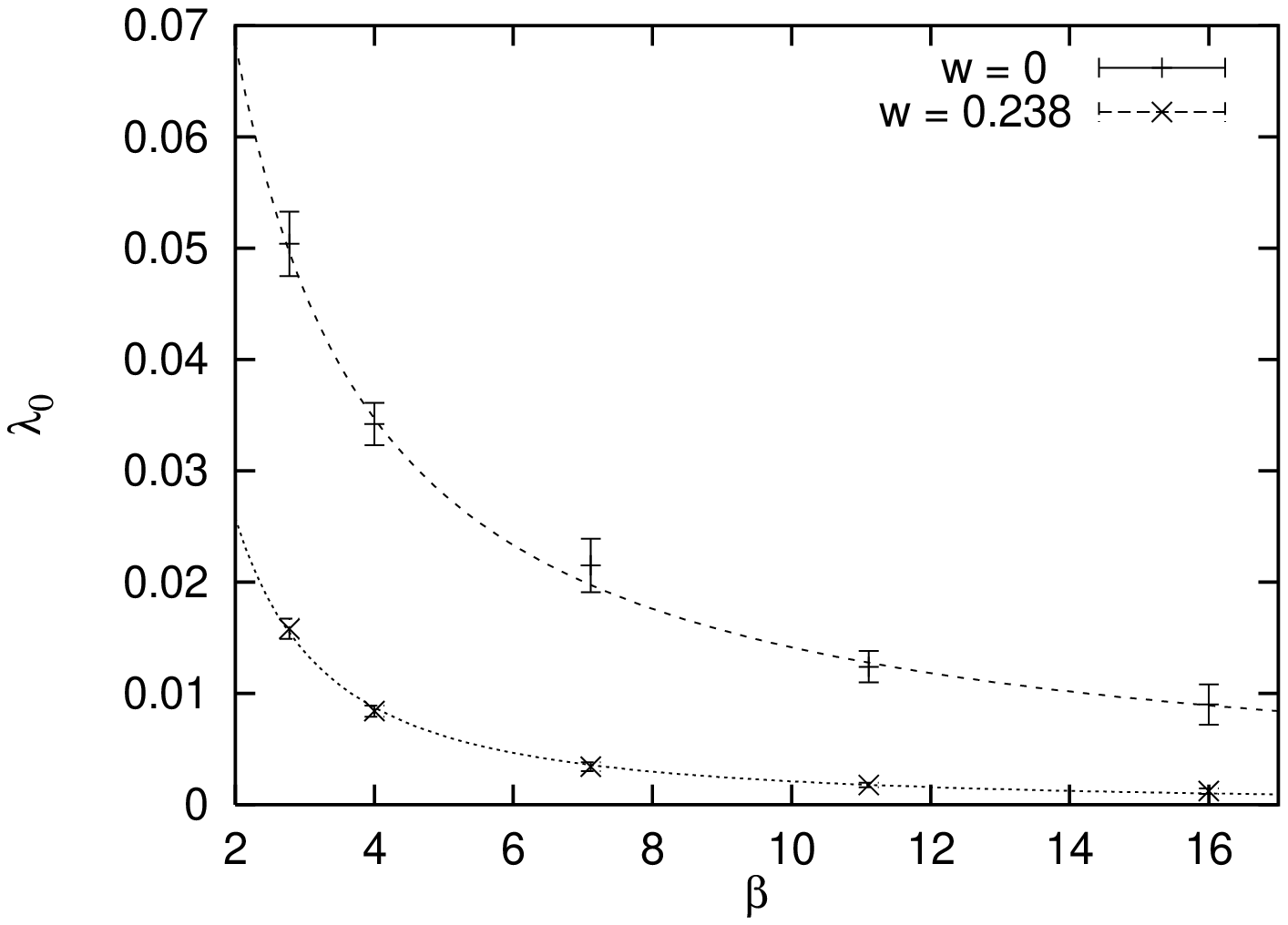,width=\figwidth,angle=270}
%\def\fpsangle{270}
%\epsfxsize=75mm
%\fpsbox{eigenval.ps}
\parbox{\textwidth}{ \caption{\label{zm-beta-fig} \sl 
The average lowest absolute eigenvalues $\lambda_0(\beta)$ for $w=0$ 
(+) and $w=0.238$ ($\times$). Physical size and coupling are fixed so that
$L^2/\beta=36$. The fits correspond to $\lambda_{0}\propto 1/\beta$
for $w=0$, and $\lambda_{0}\propto 1/\beta^{3/2}$ for $w=0.238$.
}}
\end{center}
\end{figure}

To optimize the fat link weight $w$, 
we study the behavior of the fermion matrix in a gauge field background
of topological charge $Q_{top}=1$. In the continuum limit this matrix
should have two zero eigenvalues (corresponding to the two flavors), 
and we choose $w$ such that we reproduce this property on the lattice 
as accurately as possible. 
To this end, we generate a sample of quenched configurations 
to a fixed topological 
charge and a fixed (large) value of $\beta$. 
In these configurations there are
two fermion matrix eigenvalues of particularly small absolute value.
We tune $w$ in order to minimize their average, $\lambda_0$.
This yields $w=0.238$. 
%Conversely, with this choice in the continuum limit $a\rightarrow0$, 
%$\beta = 1 / (a^2 e_c^2), \ L = L_c / a$,
%we observe an improved behavior of $\lambda_0\propto a^3$ compared to 
%$\lambda_0\propto a^2$ for $w=0$, see \fig{zm-beta-fig}. 
If we fix the physical charge and size as $e_{c}$, $L_{c}$,
so that $\beta = 1/(a \, e_{c})^{2}$, $L=L_{c}/a$, then this
choice of $w$ leads in the continuum limit to an improved behavior 
of $\lambda_{0} \propto a^{3}$, in contrast to
$\lambda_{0} \propto a^{2}$ for $w=0$, see \fig{zm-beta-fig}. 
Of course, the determination of $w$ is only approximative,
because of the use of quenched configurations.
The corresponding unquenched study should be also feasible.
However, in physical theories like QCD, it is much more expensive to tune 
the parameters with unquenched configurations.

%\clearpage
\section{The perfect pure gauge action}  \label{PureGauge}

We set the pure gauge part of the action either to the
standard Wilson plaquette action $S_W[U]$ 
-- which is perfect with respect to a somewhat complicated
RGT \cite{QuaGlu} -- or to an approximation of another 
perfect action, which is of the Villain type.
The construction of that action is closer to the PD scheme
at $m=0$, and its approximation is given by
\Eqa
 S_V[U] & = & \beta \sum_x F(x)^2, \quad e^{iF(x)} = P(x) \ , 
              \quad F(x) \in (-\pi ,\pi ] \ , \label{latticeF}\\
 P(x)   & = & U_1(x) \, U_2(x+e_1) \, U_1(x+e_2)^* \, U_2(x)^* .
\Eqb

The action
$S_{V}[U]$ can be constructed in $d=2$ by
requiring the identity of correlation functions of gauge invariant 
quantities in the
continuum and on the lattice. This parallels the treatment
of the fermions that we are using here
(we recall that in the PD scheme we block massless fermions
by a $\delta$ function RGT).
We work with the (real) phases $A_\mu(y)$ (non-compact gauge fields)
of the parallel transporters $U_\mu(y)$,
and define a plaquette phase as
%with values in $\R$ 
\eqa
    \Ft(y) \ = \ A_1(y) + A_2(y+e_1) - A_1(y+e_2) - A_2(y) \ .
\eqb
It is equal to the lattice field strength $F(x)$ from 
\eqref{latticeF} modulo $2\pi$.
In the continuum we define the phase of parallel transporters 
around plaquettes 
$[y]$ corresponding to the above lattice quantity by
\eqa
    \ft(y) \ = \ \int_{\partial [y]} a_{\mu} dx_{\mu} \ ,
\eqb
%where the continuum gauge field $a=a_\mu(x)\, dx_\mu$ is considered a
%1-form.
where $a_{\mu}$ is the continuum gauge field.  
We require the equality of the continuum and lattice two-point functions
\eqa \label{twopointG}
    \langle \Ft(y) \Ft(y') \rangle_{\subs{lattice}} 
    \ = \  \langle \ft(y) \ft(y') \rangle_{\subs{continuum}} \ .
\eqb
This leads to the definition of a perfect action
\eqa
    e^{-S[A]} \ = \ \int \Dc a_\mu \prod_y \delta( \Ft(y) - \ft(y) )
                    e^{-\beta \int dx \, f(x)^2} \ ,
\eqb
where $f(x)$ is the continuum field strength. We assumed the perfect action to 
depend on the variables $F_y$ only, i.e.\ on all closed loops except Polyakov
loops, like the continuum pure gauge action.

For the continuum action we may decompose the field strength 
$f(x)$ in averages 
$\fb(y)$ over the plaquettes $[y]$ and fluctuations $f_h(x)$ (with
$\int_{[y]} \! dx \, f_h(x) = 0$)
\eqa
 \int dx \, f(x)^2 = \sum_y \int_{[y]} dx (\fb(y)+f_h(x))^2 
                   = \sum_y \fb(y)^2 \ + \ \int dx f_h(x)^2 \ .
\eqb
Furthermore, we decompose the continuum configuration integral 
into an integral 
over all gauges, over the torons (the constant part of the gauge field 
$a_\mu(x)$, see \secref{Torons}), an integral over the field strength 
configurations $[f(x)]$, and an integral over the vortex configuration 
$[n_y]$ for each plaquette,
\eqa \label{Stokes}
    n_y = (\ft(y)-\fb(y)) / (2\pi) \ .
\eqb
The deviations from Stokes' theorem  arise for topologically 
non-trivial gauge fields. 
For the $U(1)$-topology on the 2d torus $\Tc_2$ it would be 
sufficient to allow for vortices at one specific point $x$. 
However, for a local formulation, 
we may decompose $\Tc_2$ into patches corresponding to all 
plaquettes $[y]$, see Ref.\ \cite{Luescher}.

The integral $\Dc[f(x)]$ factorizes with respect to $\fb(y)$ and $f_h(x)$,
and we are left with
\eqa
    e^{-S[A]} \ = \ \prod_y \sum_{n_y} \int d\fb(y) \
                    \delta(\Ft(y)-\fb(y)-2\pi n_y) \ e^{-\beta\fb(y)^2} \ ,
\eqb
which leads to
\eqa
 S[A]     \ = \ \sum_y s(\Ft_y) \ , \qquad
 s(\Ft_y) \ = \ -\ln \left( \sum_{n_y} e^{-\beta(\Ft(y)-2\pi n_y)^2} \right) 
          \ \simeq \ \beta F(y)^2 \ .
\eqb
Remember that $F(y)$ is the projection $\mod{2\pi}$ of $\Ft(y)$ into the 
standard interval $(-\pi,\pi ]$. 
The above approximation, $S_V$ in \eqref{latticeF}, 
neglects only terms of order $e^{-4\pi^2\beta}$, if $\Ft_y$ 
is not near an exceptional value $\pm\pi$ (which is strongly suppressed at
$\beta\geq1$).

%We add two more remarks on the pure gauge part. 
Note that the above block scheme relies on the fact 
that -- due to Stokes' theorem --
blocking the 1-form $a$ on links is consistent with blocking the 2-form $f=da$ 
on plaquettes modulo vortices, see \eqref{Stokes}. In this sense, we proceed 
in analogy to the derivation of staggered fermions from DK fermions \cite{BJ},
see also the discussion in Ref.\ \cite{Dilg}.

As a last remark on the pure gauge part, note that it is not
straightforward to follow 
the above procedure for $U(1)$ gauge theories in $d>2$.
It is easy to derive the condition for the lattice 
two-point functions corresponding to \eqref{twopointG}, but
the lattice action cannot be derived by a direct inversion 
of the two-point function,
since the plaquette variables are not independent any more.

%\clearpage
\section{An adequate Hybrid Monte Carlo algorithm} \label{HMC}

The feasibility of simulations with a large number of couplings in the action
is a major obstacle against the use of quasi-perfect actions.
%for dynamical fermions.
The computational cost of such calculations is mainly given by frequent
solutions of linear equations with the fermion matrix, which in turn need
a large number of matrix multiplications. If the fermion matrix 
${\bf m}_{xy}[U]$ is at hand, 
one matrix multiplication means $n_{\subs{coupl}}\times V$ complex
multiplications, where $n_{\subs{coupl}}$ 
is the number of path couplings to a fixed 
spinor $\bar \Psi (x)$ (or $\Psi (x)$), and $V$ is the lattice volume. 
For our truncation, one has  
$n_{\subs{coupl}}=24$ in $d=2$, compared 
to $n_{\subs{coupl}}=4$ for massless
standard staggered fermions, see \eqref{Sf}.
In spite of the relatively modest overhead,
saving the fermion matrix ($n_{\subs{coupl}}\times V$
complex numbers) --
before any solution of a linear equation --
requires a large storage space, which
can be a serious problem in $d=4$. On the other hand, any 
complicated vertex structure, i.e.\ additional parallel transporters between 
$\bar \Psi (x)$ and $\Psi (y)$ on fixed sites,
do not require additional computational costs in 
leading order, since the structure of ``hyperlinks'' remains unaltered,
see also Ref.\ \cite{Norbert}.
An example for such additional parallel transporters are the staple
terms, which we used to build the specific hyperlink called
``fat link''.

To suppress the increase of computer time needed, 
which in our case would naively amount to a factor of
$6$, we make use of the structure of the 
Hybrid Monte Carlo algorithm \cite{HMC}. 
It consists of a proposal, which is derived by a numerical integration of 
Hamiltonian equations in an artificial time (the Molecular Dynamics (MD) 
step), and a Metropolis acceptance decision. 
Detailed balance is guaranteed by a proposal symmetry within the 
MD steps, and the Metropolis decision with respect to the complete action $S$. 
The r\^{o}le of the MD proposal is to push the system 
far in configuration space 
without deviating much from the hyperplane $S = \unit{constant}$. Thus the 
acceptance probability $P_a$ in the Metropolis decision can be kept at 
$P_a \approx 0.5$ for large integration intervals in the artificial time,
and therefore large moves of the system in configuration space. 
However, we are free to choose another action $S'$ to be held approximately 
constant. This will lead to smaller acceptance probabilities, therefore the
difference of $S$ and $S'$ should better not be too large.

\begin{figure}[htb]
\begin{center}
\epsfig{file=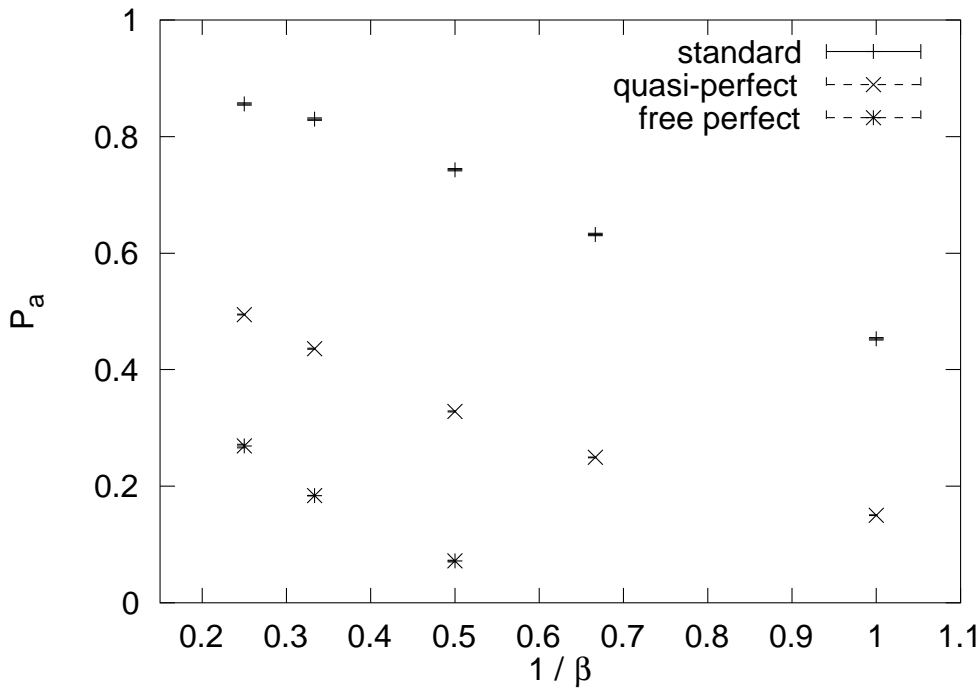,width=\figwidth,angle=270}
\parbox{\textwidth}{ \caption{\label{Pa-fig} \sl
The acceptance rate $P_{a}(\beta )$ on a $16 \times 16$ lattice.
The actions involved are:
standard staggered fermions with gauge action $S_{W}$
in the MD step and free perfect in the Metropolis step
($\times \!\!\!\!\! +$); 
standard staggered with $w=0.238$ and gauge action $S_{V}$
in the MD step and quasi-perfect in the Metropolis step ($\times$); 
standard staggered fermions and $S_{W}$ in both cases ($+$).
(The terms 'free perfect' and 'quasi-perfect' are defined in Section 8.)}}
\end{center}
\end{figure}

We choose the standard action in the MD steps, including the fat link
parallel transporters for $w \neq 0$. The gauge action is not altered.
\fig{Pa-fig} shows the behavior of the acceptance probability 
$P_a(\beta)$: it decreases as one approaches the strong coupling regime.
%%%%%%%%%%%%%%%%%%%%%%%%%%%%%%%%%%%%%%%%%%%%%%%%%%%%%%%%%%%%%%%%%%%%%%
This may be expected, since for lower $\beta$, $S$ and $S'$ really describe 
different physics, namely approximate continuum behavior for the perfect 
action $S$ and strong coupling behavior for $S'$.
With the fat link included, the strong coupling region is pushed down to lower 
values of $\beta$. Therefore we obtain useful probabilities of $P_a$ 
also for lower $\beta$-values, down to $\beta \approx 1.5$.
%$\beta $~\raisebox{-.4ex}{$\stackrel{<}{\sim}$}~1.5.

The relation of computer time needed for the MD proposal and for the Metropolis
decision is approximately given by the number of MD steps needed to achieve 
a sufficiently precise integration of the Hamiltonian equations 
in the artificial time. In our case we use a random choice between 
6 and 12 MD steps corresponding
to an artificial time interval of $0.25$ to $0.5$. 
Since the full fermion matrix
multiplication needs a factor of 6 more floating point operations, this leads
to an overall factor for the computer time of 
$\approx 1.5$ with the above method,
compared to $6$ for MD proposals using the full quasi-perfect action. 
This must be compared
with the additional autocorrelation time due to the decreased $P_a$, see
\fig{Pa-fig}, which shows that there is still a substantial gain left.
We find it worthwhile to check, whether it is possible to proceed
in a similar way in $d=4$ models. This will depend on the much stronger 
increase in the number of couplings for quasi-perfect actions, on the 
number of MD steps in these models, and on the behavior of $P_a$. 

There are other effects, which can increase the autocorrelation time when one
approaches the continuum physics with improved actions. Particularly
in the Schwinger model,
the approximate zero-modes will become sharper, and for any observable
with zero-mode contributions a special treatment becomes necessary.
For instance, one may measure the contributions from different topological 
sectors separately, and determine their relative contribution after a 
re-weighting of the sectors in question (in the sense of a discrete 
multi-canonical simulation \cite{Neuhaus}), as outlined in Ref.\ \cite{Diss}.
Alternatively, one could use a Polynomial Hybrid Monte Carlo
algorithm \cite{PHMC}, recently proposed
to deal with zero-mode contributions.
In both cases, the problem of effectively changing the topological sector 
remains. For our case, we used instanton hits matched to the present 
topological zero-modes, see Ref.\ \cite{Zeromodes}.

We want to mention another two numerical problems. First, the
calculation of connected contributions (see Section 9)
requires the calculation of
propagators (${\bf m}^{-1}_{xy}$) 
with sources at all lattice points $x$. Doing this 
for each configuration would need a multiple of the time needed to generate 
these configurations. 
One often uses methods like the noisy estimator to circumvent
this problem. In our case, however, we decided to really calculate all 
propagators, but not for all configurations, since there are large 
autocorrelation times in the fermionic observables (probably due to topological
quantities with large autocorrelation times). This makes it also possible
to average over the propagator source for disconnected contributions, and 
thereby decrease the statistical fluctuations of the single measurements.  

Secondly, it is difficult to evaluate small masses from 
correlations with large relative errors. 
For artificial data of a decay $c(t)=c_0\cosh(m(t-L/2))$,
modified with uncorrelated Gaussian 
fluctuations $\delta c(t)/c(t) = \epsilon$, 
$m\leq\epsilon$, we found our fitting procedure to yield results for 
$m$ in a single $\cosh$-fit, which were not resolved from zero despite of error
bars of order $10^{-7}$. 
We therefore trust in decay masses only, if they are resolved from zero
by their error bars. This is a problem for the evaluation of the $\pi$-mass,
where similar effects appear, in particular for small fit intervals.
%see below.   wo?

Altogether we generated $\approx 3\cdot 10^5$ configurations with maximal
autocorrelation times up to $100$ for topological observables, and up to $25$ 
for other gauge field observables. For fermionic observables we measured after
steps of $100$ configurations. 
Even after that reduction we found autocorrelation times up to 3.  

%\clearpage
\section{Results for the torons} \label{Torons}

Here we present the results for the free truncated
perfect action for staggered fermions
\begin{itemize}
\item with the minimal vertex ($w=0$) (``{\em free perfect}''), 
together with the standard pure gauge plaquette
action $S_W$, 
\item and for our extended approximation
of a perfect action ({\em quasi-perfect}: $w_{FL}=0.238$,
gauge action $S_V$).
\end{itemize}

First, we investigate the behavior of the toron part of the gauge
field, i.e.\ its constant mode on the 2d continuum torus
\footnote{
For a lattice version of the average in \eqref{t-cont}, 
the $2\pi$-ambiguity of
the phases $A_\mu(x)$ of the link variables $U_\mu(x)$ have to be resolved by 
fixing the gauge invariant lattice field strength $F(x)\in (-\pi,\pi ]$, 
see \eqref{latticeF}, and by the requirement 
$F(x)=A_1(x)+A_2(x+e_1)-A_1(x+e_2)-A_2(x)$, see Refs.\ \cite{Diss,Fx}.
%%%$\Tc$ of size $L_1 \times L_2$%
}
\eqa \label{t-cont}
    t_\mu = \frac{1}{L_1 L_2} \int dx \ a_\mu(x) \ .
\eqb 
It is convenient to define a dimensionless quantity 
$u_\mu=(e L_\mu/2\pi)\,t_\mu$.
Under large gauge transformations it transforms as $u_\mu \rightarrow 
u_\mu+n_\mu,\ n_\mu \in \Z$. Its effective fermion-induced action $\Gamma(u)$ 
(the $\half \log$ of the fermion determinant) decouples 
from all other parts of 
the gauge field, and is independent of the gauge coupling $e$,
\eqa
\Gamma(u) \ = \ 2 \ln \vert \theta_1 \left( u_2 + i \tau u_1 \right ) \vert
          \ - \ 2 \pi \tau \, u_1^2 \ , \quad \tau = L_2/L_1 \ ,
\eqb
where $\theta_1$ is the Jacobian $\theta$-function.
For details of the calculation in the two-flavor 
case we refer to Ref.\ \cite{AJ}.

We therefore expect an approximately perfect behavior of toron-dependent
observables already with the free perfect fermion. We measured the Fourier 
coefficients $c_{n_1,n_2}$ of the toron distribution
\eqa
 c_{n_1,n_2} \ \equiv \int d^2u \ e^{\half \Gamma(u)} \ 
 \cos( 2\pi u_1 n_1) \ \cos( 2\pi u_2 n_2) \ ,
\eqb
and results for various $\beta$ and lattice sizes are given 
in \tab{toron-tab}.
\begin{table}[htb]
\begin{center}
\begin{tabular}{| l || l | l | l | l |}
\hline
 6$\times$16-lattice, $\beta\!=\!10$     
 & $c_{1,0}$ & $c_{2,0}$ & $c_{0,1}$ & $c_{0,2} $ \\
\hline
standard 
 & $-0.7619(9)$ & $0.3415(16)$ & $-0.0061(15)$ & $0.0014(11)$ \\
free perfect 
 & $-0.7441(7)$ & $0.3084(15)$ & $-0.0079(11)$ & $0.0009(8)$  \\
quasi-perfect 
 & \ \ ---  & \ \ ---  & \ \ ---  & \ \ ---  \\
continuum 
 & $-0.7435$  & $0.3079$  & $-0.0075$  & $0.0002$  \\
\hline
\hline
 16$\times$16-lattice, $\beta\!=\!3$     
 & $c_{1,0}$ & $c_{2,0}$ & $c_{0,1}$ & $c_{0,2} $ \\
\hline
standard 
 & $-0.299(8)$  & $0.047(5)$ & $-0.290(7)$  & $0.040(5)$ \\
free perfect 
 & $-0.301(10)$ & $0.038(8)$ & $-0.312(10)$ & $0.050(6)$ \\
quasi-perfect 
 & $-0.322(9)$  & $0.052(6)$ & $-0.322(9)$  & $0.038(7)$ \\
continuum
 & $-0.322$ & $0.043$ & $-0.322$ & $0.043$ \\
\hline
\end{tabular}
\parbox{\textwidth}{ \caption{\label{toron-tab} \sl
Some toron expectation values (defined in Eq.\ (8.3))
in the standard, free perfect, and 
quasi-perfect case, compared with the continuum results.}}
\end{center}
\end{table}
%%%%%%%%%%%%%%%%%%%%%%%%%%%%%%%%%%%%%%%%%%%%%%%%%%%%%%%%%%%%%%%%%%%%%%%%
There is no dramatic deviation from the continuum results 
within small errors, even with the vertex based on shortest paths only.
This is clearly in contrast to the standard staggered fermions.

%\clearpage
\section{The ``meson'' spectrum} \label{Results}

\subsection{Observables}

Finally, the ``meson'' spectrum is derived from the 
correlation functions of point-like
and one-link lattice bilinears corresponding to the isovector 
and isoscalar current (see Ref.\ \cite{Diss}),
\Eqa
 \mbox{isovector:} \qquad M^0_{+-}(x) & = & 
(-1)^{x_2}\ \bar \Psi (x) \Psi (x)\ , \nonumber \\
 \mbox{isovector:} \qquad M^{1-}_{--}(x) & = & (-1)^{x_1+x_2} \ \nonumber
 [\bar \Psi (x) \Psi(x+e_1) \, - \, \bar \Psi (x+e_1) \Psi (x)] \ ,\\
    \mbox{isoscalar:} \qquad M^{1+}_{++}(x) & = & 
    \bar \Psi (x) \Psi (x+e_1) \, + \, \bar \Psi (x+e_1) \Psi (x)] \ .
\Eqb
The isospin $I=\pm 1/2$ labels the two flavors of Dirac fermions 
corresponding to staggered fermions for $d=2$. From the bilinears 
$M^\alpha(x)_{\sigma_1\sigma_2}$, $\alpha=0, \, 1\!- , \, 1+,$ 
we derive correlation functions at fixed spatial momentum 
$p_1$ in the usual manner,
\eqa
    c^\alpha_{\sigma_1\sigma_2}(t;p_1) \ = \ \sum_x e^{i p_1 x_1} \ \langle \
    M^\alpha_{\sigma_1\sigma_2}(x_1,t) \ M^\alpha_{\sigma_1\sigma_2}(0,0) 
    \ \rangle \ .
\eqb
Here, it is helpful to consider the mechanisms
which determine the current 
correlation functions in the continuum. There are three possible contributions 
to fermionic two-point functions in the 2-flavor Schwinger model 
\cite{Sachs,AJ}.
The connected part (with propagators from $x$ to $y$ and back) is
the only non-vanishing contribution for the isovector current correlation.
It decouples from the gauge field, except from its constant (toron) part. 
The corresponding intermediate particle is massless (we call it ``pion''). 
For the isoscalar current correlation, the 
disconnected part (propagators from $x$ to $x$ and from $y$ to $y$) also 
contributes.
It can be treated by gauge invariant point splitting, which introduces an 
(anomalous) coupling to the gauge field. Combined with the disconnected part, 
the correlation is then determined by an intermediate massive particle (the 
``$\eta$-particle'').

The current correlation functions receive only contributions from gauge fields 
with zero topological charge. Gauge configurations in non-trivial sectors
are suppressed by topological zero-modes, and the currents do not couple to
the poles in the propagators which arise from those zero eigenvalues. In
contrast, the correlation functions of scalar and pseudoscalar densities
pick up zero-mode contributions, 
which arise from the cancellation of these poles with 
the zeros of the fermion determinant. For a discussion on how these mechanisms 
are realized on the lattice with staggered fermions we refer to 
Ref.\ \cite{Diss}.

In a given time slice we cannot fix the quantum number
$\sigma_2$. Thus the corresponding 
correlation functions obtain contributions from $\sigma_2=\pm$,
i.e.\ with and without a sign factor $(-1)^{x_2}$. 
In this way, current and density correlations are always intertwined.
The latter are sensitive to zero-mode contributions from non-zero topological
charges. To some extent, it is possible to project out these contributions, 
using
\eqa
    C^\alpha_{\sigma_1\sigma_2}(t;p_1) \ = \ 
    c^\alpha_{\sigma_1\sigma_2}(t;p_1) + \half \sigma_2 \ 
    [ \ c^\alpha_{\sigma_1\sigma_2}(t+1;p_1)  
    +   c^\alpha_{\sigma_1\sigma_2}(t-1;p_1) \ ] \ .    
\eqb
Of course, this projection is only good at small energies in the undesired 
channel. However, we seem to reduce the error in the correlation functions 
in this way by large factors of about 20 and more. 
This may be in part due to the suppression of the zero-mode contributions to 
scalar-density-like correlations, which have particularly 
high statistical errors (autocorrelation times). 
For the correlations $C^\alpha_{\sigma_1\sigma_2}(t;p_1)$ we obtain sensible 
fits with the ansatz
\eqa
    C^\alpha_{\sigma_1\sigma_2}(t;p_1) \ = \ \sum_{\sigma_2=\pm}
    const_{\sigma_2} \cosh[ \, E_{\sigma_2} \, (t - T/2) \, ] \ .       
\eqb
The fit interval is chosen so that $C^\alpha_{\sigma_1\sigma_2}(t=0;p_1)$ 
does not contribute, except for the correlation corresponding to the isoscalar 
current ($\eta$-particle). There the fits are particularly stable with respect 
to the size of the interval size. 
In this way we always achieve $\chi^2/d.o.f. \leq 2$.

\subsection{Results}

We now compare results for the $\pi$ and $\eta$ spectrum
obtained from standard staggered fermions and from our
quasi-perfect staggered fermion action.

For any action based on the truncated perfect free fermion,
we expect the dispersion relation
of the decay energies $E(p_1)$ of time slice correlations with fixed spatial 
momentum $p_1$ to approximate the continuum form $E(p_1)^2 = m^2 + p_1^2$. 
Here, we consider the $\eta$-particle, given by the $M^{1+}_{++}$ lattice 
field, and the $\pi$-particle, given by the $M^{1-}_{--}$ field without 
disconnected contributions.
The latter choice is motivated by the requirement that masses 
should be resolved
from $0$ within their error bars, see \secref{HMC}. 
It relies on the absence of disconnected
contributions to the isovector channel in the continuum. Indeed, for low 
$\beta$-values, i.e. $\pi$-masses further away from zero,
it appears that the 
masses evaluated from $M^{1-}_{--}$ including disconnected contributions, 
as well as masses from $M^0_{+-}$, are significantly smaller.
Thus we take the worst case of a $\pi$-mass with respect to the continuum
massless behavior, and the easiest case for a comparison of the different
improvement steps.
In Figures 12 and 13 we give the energies for the 
$\pi$ and the $\eta$, depending on $p_1$ for $\beta=3$ 
with the standard staggered 
action and with the quasi-perfect action.
These and the following results are obtained on a $16 \times 16$ lattice
(where we have momenta $p_1=n\pi /8,\ n=0,\dots ,4$).
The lines show the continuum dispersion relation with the mass derived from 
$p_1=\pi/8$. In the quasi-perfect case they fit the data in an excellent 
way.
\begin{figure}[htb]
\begin{center}
\epsfig{file=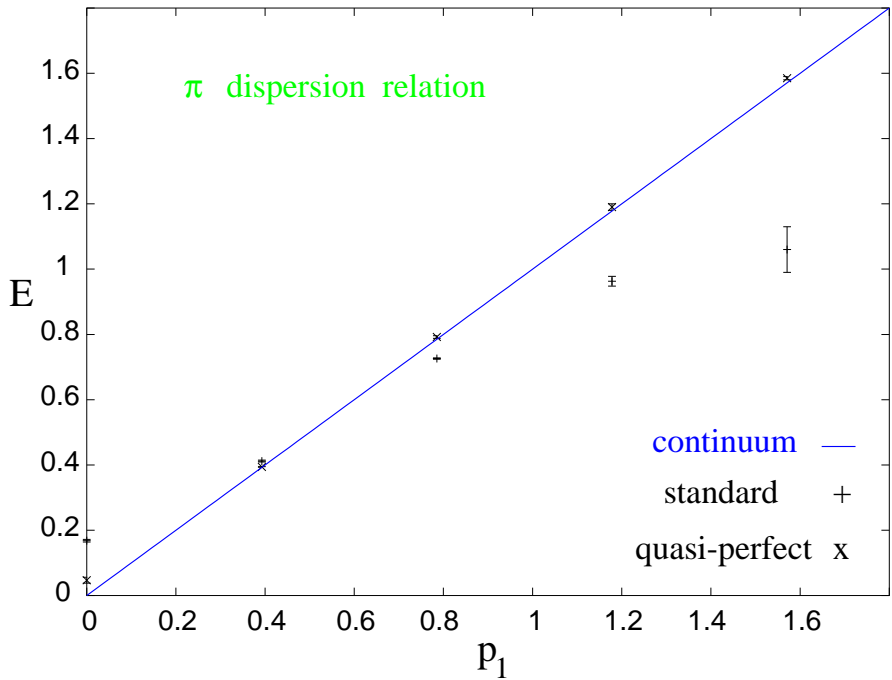,width=\figwidth,angle=270}
\parbox{\textwidth}{ \caption{\label{disp-pi} \sl 
The dispersion relation for the pion with standard and
the quasi-perfect action at $\beta =3$.}}
\end{center}
\end{figure}
\begin{figure}[htb]
\begin{center}
\epsfig{file=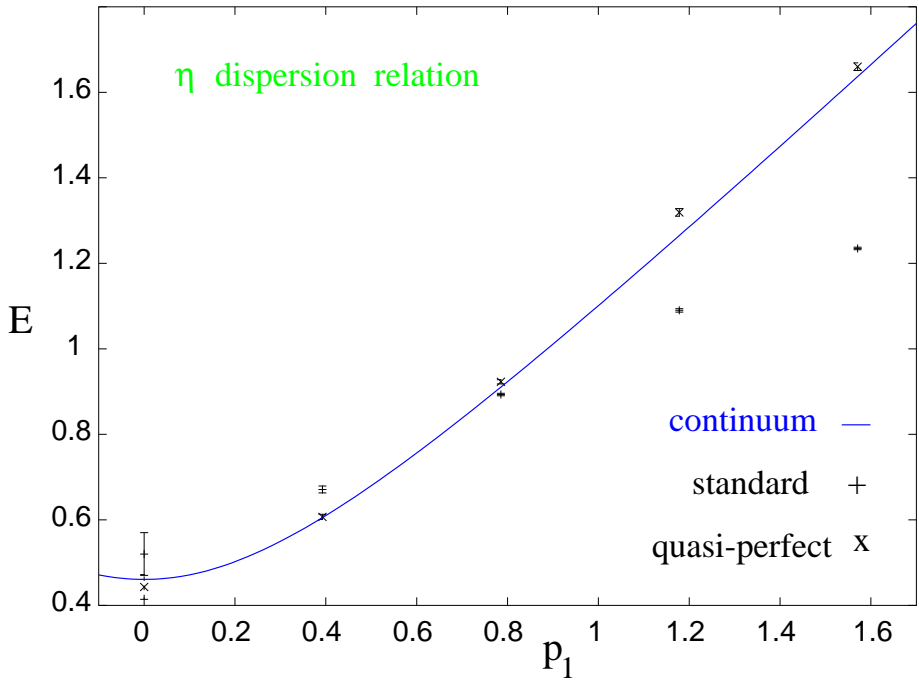,width=\figwidth,angle=270}
\parbox{\textwidth}{ \caption{\label{disp-eta} \sl 
The dispersion relation for the $\eta$-particle with 
standard and the quasi-perfect action at $\beta=3$.}}
\end{center}
\end{figure}

Finally, we show the $\beta$-dependence of the $\pi$- and $\eta$-mass in
Figures 14 and 15. In the Schwinger model, asymptotic scaling 
for the $\eta$-mass is given by $m_\eta^2=2/\pi\beta$.
\begin{figure}[htb]
\begin{center}
\epsfig{file=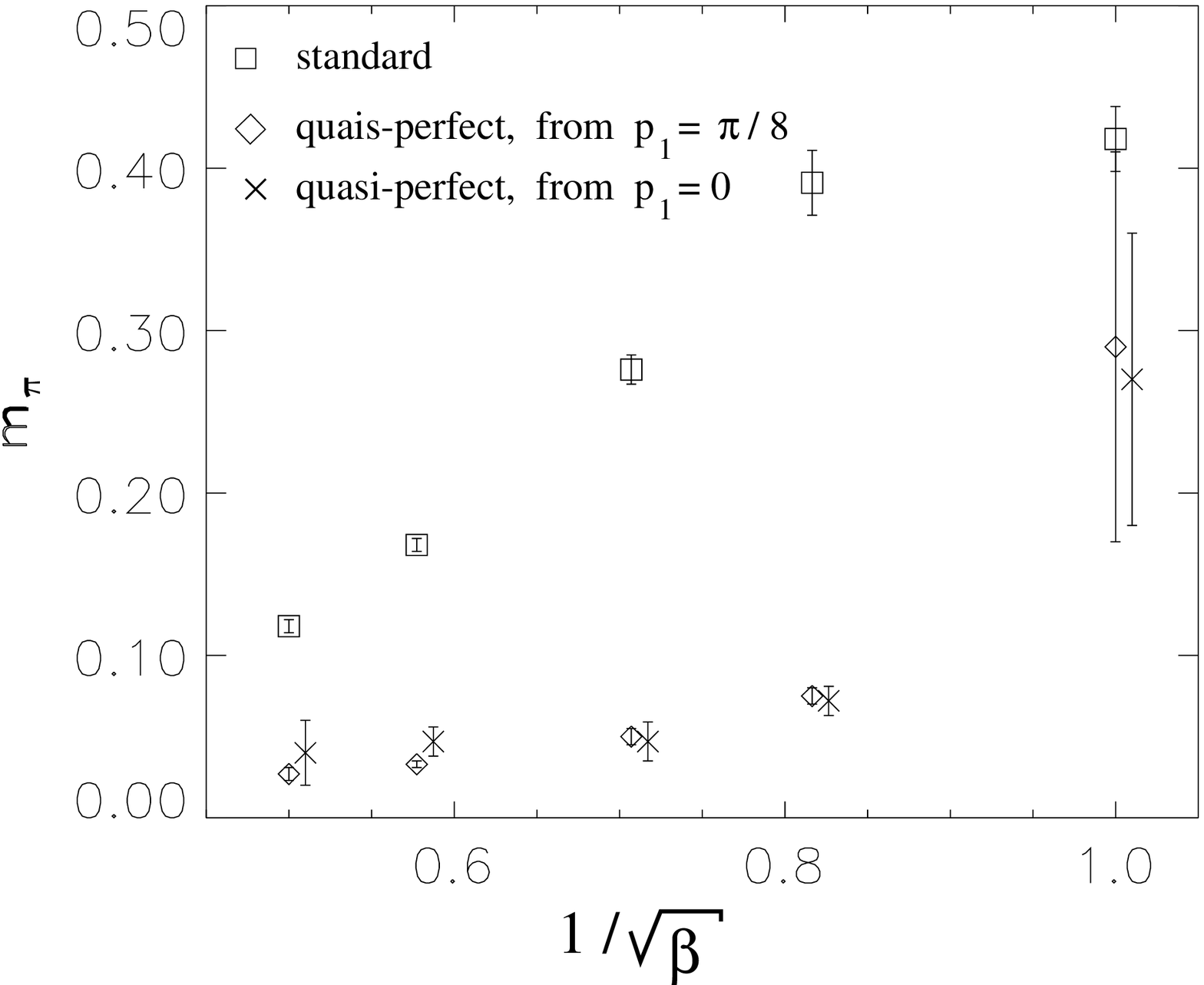,width=\figwidth}
\parbox{\textwidth}{ \caption{\label{pi-beta} \sl 
The $\beta$-dependence of $m_\pi$ with standard staggered fermions (squares) 
and quasi-perfect fermions, derived from $p_1=0$ (crosses) and $p_1=\pi/8$ 
(diamonds).}}
\end{center}
\end{figure}
\begin{figure}[htb]
\begin{center}
\epsfig{file=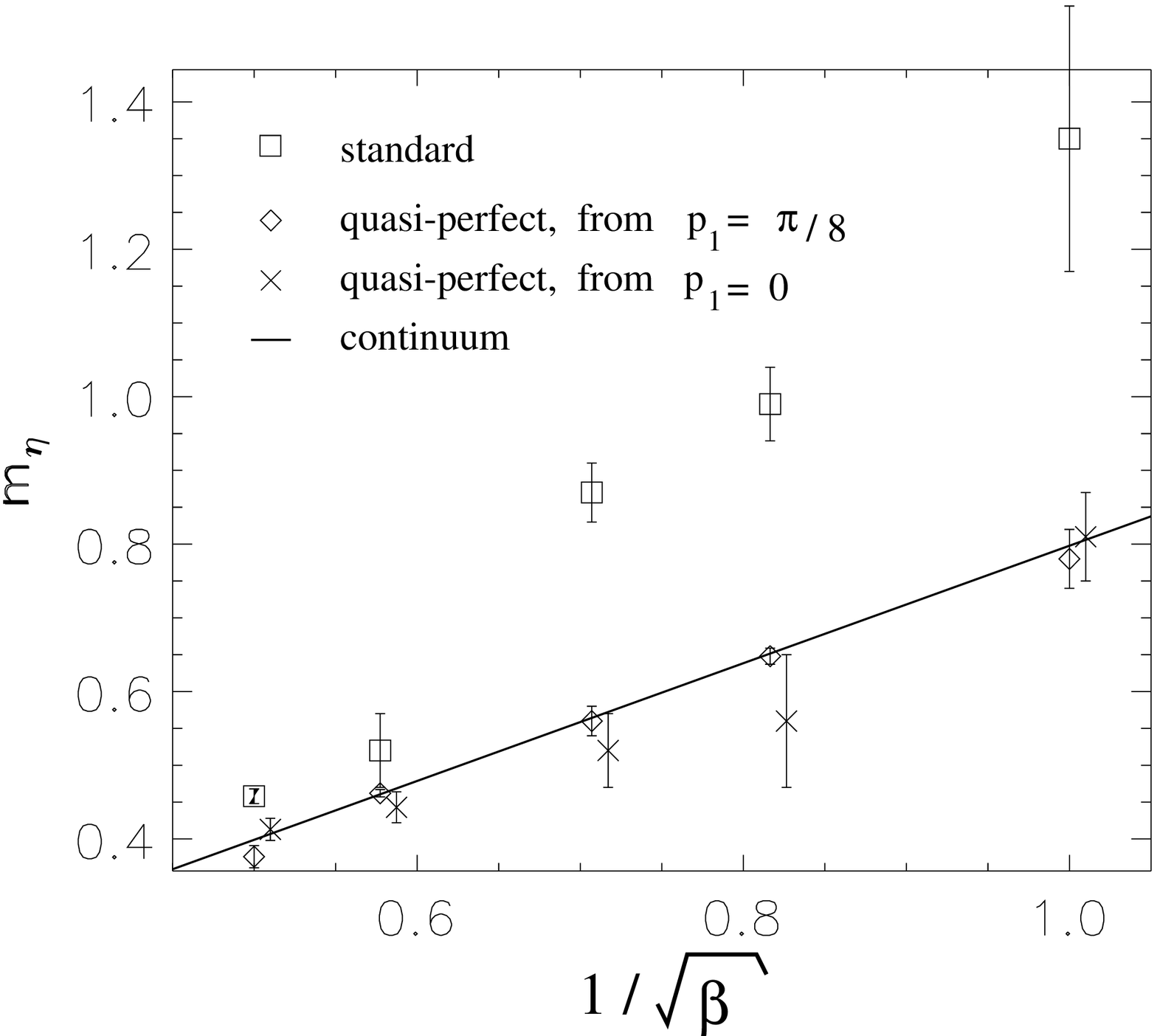,width=\figwidth}
\parbox{\textwidth}{ \caption{\label{eta-beta} \sl 
The $\beta$-dependence of $m_\eta$ with standard staggered fermions (squares) 
and quasi-perfect fermions, derived from $p_1=0$ (crosses) and $p_1=\pi/8$ 
(diamonds).}}
\end{center}
\end{figure}
The nearly perfect dispersion relation of the quasi-perfect action makes 
it possible to derive the masses in a
straightforward manner also from higher momenta, which leads to smaller errors
for our data.

\begin{figure}[htb]
\begin{center}
\epsfig{file=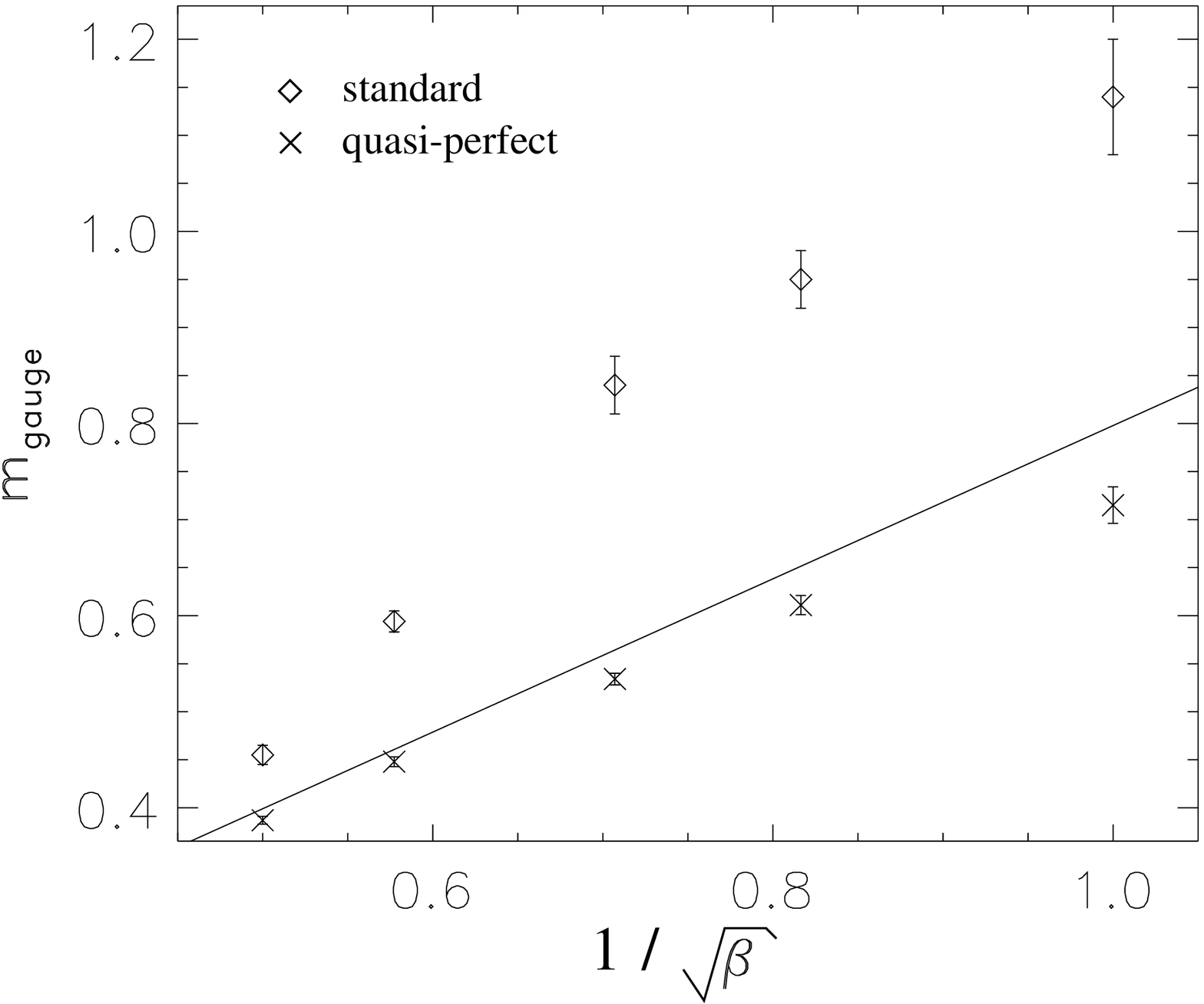,width=\figwidth}
\parbox{\textwidth}{ \caption{\label{plaq} \sl 
The $\beta$-dependence of $m_{gauge}$ with standard staggered fermions 
(diamonds) and quasi-perfect fermions (crosses).}}
\end{center}
\end{figure}

The $\eta$-mass should be equal to the decay mass $m_{gauge}$ of the 
plaquette correlation for zero spatial momentum. 
In our case, these are the most precise mass values. 
They can be extracted from a simple $\cosh$-fit.
However, for reasons that we do not fully understand,
$m_{gauge}$ does not follow the asymptotic scaling 
prediction as closely as $m_{\eta}$, see \fig{plaq}.

%\clearpage
\section{Conclusions}

Spectral and thermodynamic results for free, truncated perfect staggered
fermions have been presented before \cite{BBCW}, but the improvement
did not really reach a satisfactory level there. We now
pushed that improvement significantly further, mainly thanks to
the new blocking scheme, which we call {\em partial decimation}, but 
also with the help of a new truncation technique ({\em mixed periodic}
boundary conditions). We now reached a level of excellent
improvement, similar to the results for truncated perfect
Wilson-type fermions.

We extended the construction of perfect actions
to non-degenerate flavors, and we could preserve the same level
of improvement after truncation also in that case.
This is potentially important for the study of the decoupling of 
heavy flavors, or for QCD simulations with realistic quark masses.
\footnote{However, in that case the adequate coupling to gauge fields
still has to be worked out.}

At $m=0$, which is well described by
staggered fermions, the PD scheme is optimally local if we just use
a $\delta$ function RGT. This simplifies the relation
to the continuum $n$-point functions, and numerical RGT steps,
which could be performed for interacting theories (in the classical
limit).

In $d=2$, we added an Abelian gauge field to
this fermion along (selected) shortest lattice paths
plus a ``fat link''. In the latter, the weight $w$ of the staple term
is optimized effectively, by minimizing the lowest eigenvalues of the
fermion matrix, which led to $w=0.238$.

To simulate the 2-flavor Schwinger model,
we applied a variant of the Hybrid Monte Carlo algorithm. It uses
a simplified action in the identification of possible molecular
dynamics steps, and the quasi-perfect action in the acceptance
decision. In this way, we could avoid an increase in the computation 
effort proportional to the number of couplings.

As a vertex independent quantity, we measured the Fourier coefficients 
of the toron distribution, and we saw that even the free perfect action
(without fat link) helps to move substantially closer to the
continuum values. 

Next we considered the dispersion relation of the ``$\pi$'' and 
``$\eta$'' particle, and we extracted the masses
$m_{\pi}$ and $m_{\eta}$ at varying $\beta$. We found a tiny
pion mass down to 
$\beta $~\raisebox{-.4ex}{$\stackrel{<}{\sim}$}~1.5
which confirms an excellent scaling
behavior of our quasi-perfect action, while $m_{\eta}(\beta )$ follows
the asymptotic scaling very closely.
A better scaling is the actual goal of the construction
of improved actions, but it was observed before in other models
that quasi-perfect actions also tend to improve the asymptotic scaling
\cite{GN,gauge1}.

These results, in particular the scaling 
%-- which is the actual purpose of the improvement of the action --
is even better than the one obtained by using 
Wilson-type fermions with a classically perfect vertex
function (parameterized by 123 independent couplings) \cite{Lang}.
This shows that truncated perfect free staggered fermions, which are
suitably ``gauged by hand'', {\em can represent a highly improved,
short-ranged action}. 
\footnote{Similar attempts for Wilson-type fermions have not been
very successful so far. However, a new approach, which ``gauges by hand''
so that the violation of the violation of the Ginsparg-Wilson relation 
is minimized \cite{GW}, yields promising preliminary results.}
This variant of a quasi-perfect improvement program is applicable 
and promising for QCD.

\end{document}